# Superconductivity near quarter- and half-filling of a strongly correlated triangular Hubbard band in twisted trilayer $WSe_2$

Yungi Jeong[1], Kenji Watanabe[2], Takashi Taniguchi[3], Joonho Jang[1*]

[1] Department of Physics and Astronomy, and Institute of Applied Physics, Seoul National University, Seoul 08826, Korea

[2] Research Center for Electronic and Optical Materials, National Institute for Materials Science, 1-1 Namiki, Tsukuba 305-0044, Japan

[3] Research Center for Materials Nanoarchitectonics, National Institute for Materials Science, 1-1 Namiki, Tsukuba 305-0044, Japan

Correspondence to: joonho.jang@snu.ac.kr

**The triangular-lattice Hubbard model is one of the paradigmatic settings for studying correlated electronic systems with geometric frustration that can give rise to a variety of phases. Recently, twisted bilayer TMD systems have emerged as attractive experimental platforms for realising such a model in the moderate-correlation regime. Alternating twisted trilayer TMD systems are expected to realise stronger correlations than bilayer systems owing to their mirror symmetry. Here we report the discovery of superconductivity near quarter- and half-filling, together with various correlated phases, in alternating twisted trilayer $WSe_2$ ($TTWSe_2$). At half filling, a robust correlated insulating state persists over a broad range of displacement fields, signalling the strongly correlated regime. Upon doping, superconductivity emerges over an extended region of the phase diagram. Near quarter filling, another superconducting state appears flanked by correlated metals. Our results establish the $TTWSe_2$ system as a compelling platform for studying strongly correlated triangular-lattice Hubbard physics.**

Moiré superlattices based on van der Waals materials provide a highly tunable platform for studying correlated and topological electronic phases in engineered lattice potentials[1,2]. In particular, twisted or lattice-mismatched TMD moiré systems possess isolated narrow energy bands, whose bandwidth and interactions can be tuned by twist angle $\theta$ and displacement field $D$[1–4], making them promising platforms for simulating the Hubbard model physics[1,3,5]. Previous studies on twisted bilayer $WSe_2$ ($tWSe_2$) have shown that it can realise a multi-orbital Hubbard model[4,6] and that superconductivity emerges near Fermi-surface reconstruction[7–10] in the moderate-interaction regime[9,10]. These results highlight twisted TMD moiré systems as a fertile setting for studying exotic phases intertwined in interacting triangular-lattice models.

Alternating twisted trilayer TMDs are expected to provide a more favourable platform for strong correlation and tunable band structure. Because two moiré patterns are formed at the top-middle and middle-bottom interfaces with mirror symmetry, the moiré potentials can add constructively in the middle layer, generating a deep triangular quantum-well-like landscape and substantially enhancing the confinement of low-energy carriers towards the middle layer[11]. Theoretical studies on alternating twisted TMD trilayer predict an isolated moiré flat band which can be described as a triangular-lattice Hubbard model[11,12]. Taken together, these results suggest alternating twisted trilayer TMDs as a promising setup for accessing an unexplored regime of strong correlation in moiré-based triangular-lattice Hubbard systems.

In this work, we present a comprehensive electrical transport study of alternating twisted trilayer $WSe_2$ with a twist angle of 3.87°. We find a robust correlated insulating state at half filling over a wide range of displacement fields. Upon doping this insulating state, superconductivity emerges in a broad region of the phase diagram. In addition, another superconducting phase develops near quarter filling in the vicinity of correlated insulators and metals. The rich phase diagram, composed of a plethora of emergent phases and multiple superconducting phases, suggests that the $TTWSe_2$ system provides a unique opportunity to explore many-body phenomena in a strongly correlated triangular Hubbard model experimentally.

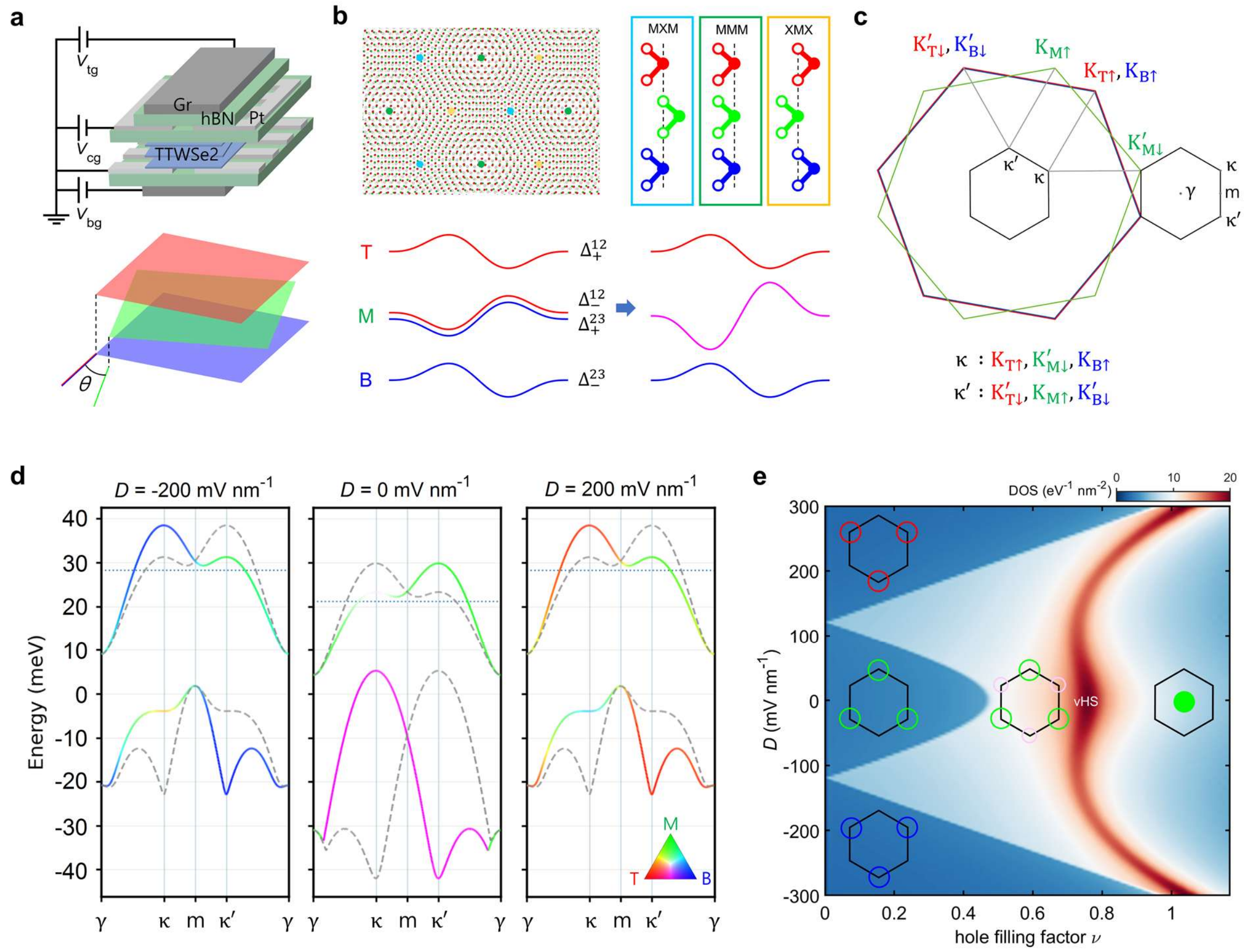

**Fig. 1 | Electronic structure of alternating twisted trilayer $WSe_2$.**
**a**, Schematic illustration of the dual-gated $TTWSe_2$ device and stacking geometry. The top and bottom $WSe_2$ layers are aligned, while the middle layer is rotated by $\theta$. **b,** Real-space structure of $TTWSe_2$. Coloured dots mark representative local atomic registries, shown at right for the MXM, MMM and XMX configurations (M = W, X = Se). Each interface generates a moiré potential and contributions from each interface combine to produce an approximately doubled moiré modulation at the middle layer (bottom). **c,** Schematic illustration of momentum-space structure. Large hexagons indicate the Brillouin zone of the top (red), middle (green) and bottom (blue) layers. Small black hexagons indicate the moiré Brillouin zone. Spin-valley-locked states $K_{T\uparrow}$, $K'_{M\downarrow}$ and $K_{B\uparrow}$ fold to κ, whereas the opposite-valley partners $K'_{T\downarrow}$, $K_{M\uparrow}$ and $K'_{B\downarrow}$ fold to κ′. **d,** Continuum-model calculation of moiré valence bands of 3.9° $TTWSe_2$ at $D$ = -200, 0, and 200 mV nm$^{-1}$. K-valley bands are shown as solid curves and colour-coded by layer polarisation, as indicated by the triangle, whereas K′-valley bands are shown in grey dashed lines. Horizontal dotted lines denote the chemical potential at half filling. **e,** Electronic DOS calculated from the continuum model as a function of hole filling factor $\nu$ and vertical displacement field $D$. Hexagons indicate Fermi-surface configurations of the K-valley band in each region separated by DOS steps and vHS. Open and filled circles denote hole and electron pockets, respectively, and the colour of each circle indicates representative layer polarisation.

**Electronic structure of $TTWSe_2$**

To understand the low-energy electronic structure of $TTWSe_2$, we extended the continuum description previously developed for twisted bilayer $WSe_2$[13] to the alternating trilayer geometry[11] (**Methods**). Due to the strong spin-orbit coupling in monolayer $WSe_2$, spin and valley are effectively locked on the energy scale considered here: holes in the K (K′) valley have spin up (down). **Fig. 1c** shows the reciprocal space structure of $TTWSe_2$. The $K_\uparrow$ ($K'_\downarrow$) valleys of the top and bottom layers are folded to the κ (κ′) point of the moiré Brillouin zone, whereas the $K_\uparrow$ ($K'_\downarrow$) valley of the middle layer is folded to κ′ (κ).

In this system, the total structural energy depends on lattice translation and is minimised when the two independently generated moiré patterns, top-middle and middle-bottom, are aligned to form a mirror symmetry about the middle layer[11] as shown in **Fig. 1b**. Importantly, owing to this mirror symmetry, the moiré potential contributions from both interfaces add constructively in the middle layer, producing a stronger effective moiré potential. As a result, electrons in the middle layer experience a deeper quantum-well-like potential landscape, raising the energy of middle-layer states relative to those of the top and bottom layers. Meanwhile, the top- and bottom-layer states, which overlap in momentum space, split into bonding-antibonding states via interlayer tunnelling (**Extended Data Fig. 3a**).

**Fig. 1d** shows the calculated continuum-model band structure in the moiré Brillouin zone, with the layer polarisation of each eigenstate indicated by the colour scale (see **Methods** and **Extended Data Fig. 3**). For the topmost K-valley band, the energy at the κ point is minimised at $D = 0$. As a finite $D$-field is applied, the energy near the κ point increases, and the states near the κ point gradually become layer-polarised; the dominant layer characteristics are determined by the direction of the displacement field. **Fig. 1e** shows the calculated density of states (DOS) as a function of moiré hole filling factor $\nu$ and displacement field $D$. Since the energies at κ and κ′ are balanced at $D = \pm 120$ mV $nm^{-1}$, resulting in a band structure similar to that of $tWSe_2$ at $D = 0$, the DOS map exhibits a characteristic W-shaped feature lying sideways where the two copies of the bilayer-like structure join at $D = 0$. The avoided crossing between top- and bottom-derived states forms a similar ω-shaped van Hove singularity (vHS) near $\nu = 0.75$.

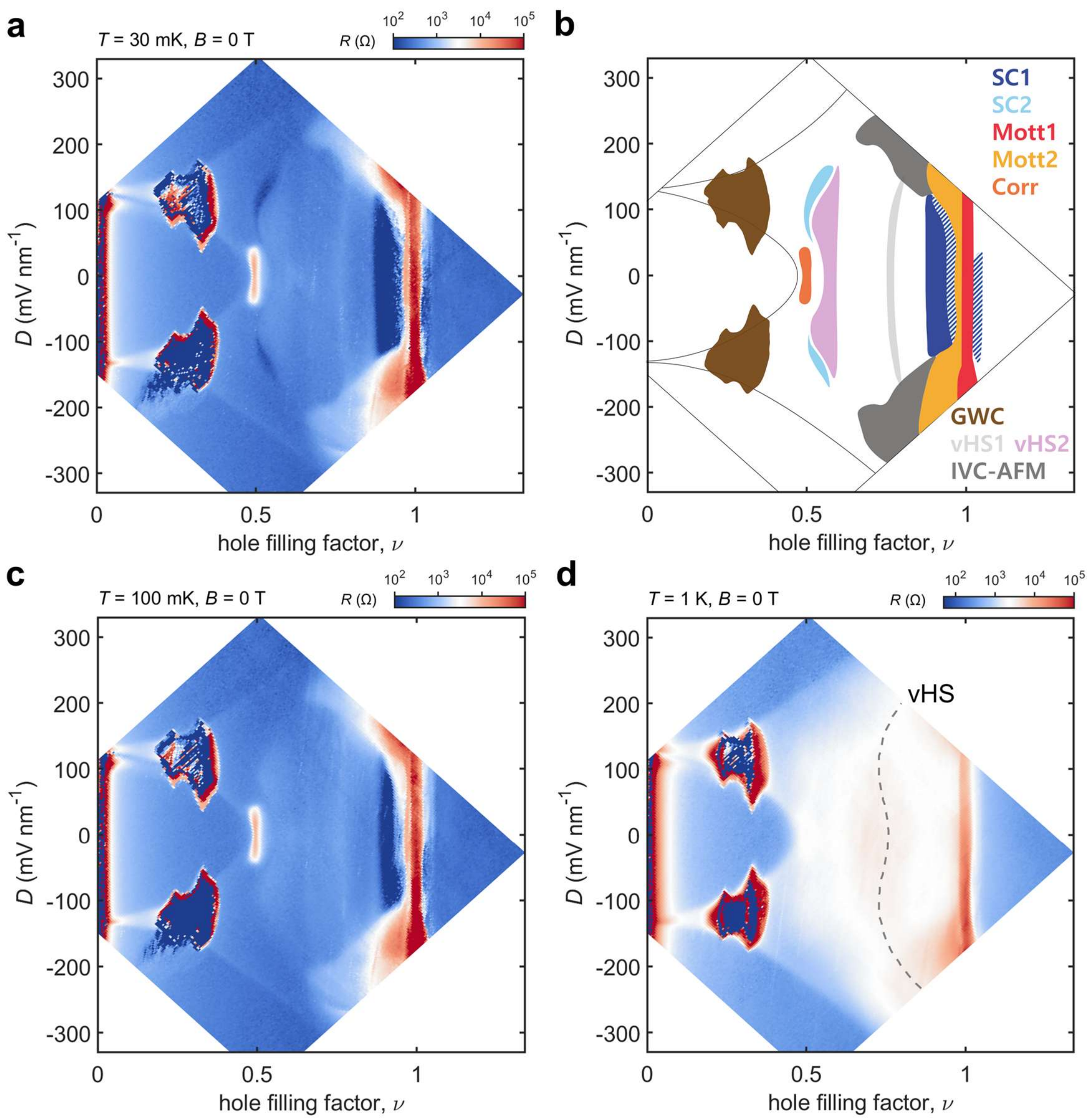

**Fig. 2 | Correlated electronic states.**

**a,c,d,** Longitudinal resistance $R$ as a function of hole filling factor $\nu$ and vertical displacement field $D$, measured at $B = 0$ T and $T = 30$ mK (**a**), 100 mK (**c**), and 1 K (**d**). **b**, Schematic phase diagram summarising the correlated states identified in **a**. Black curves separate the layer-polarised and layer-hybridised regimes. The coloured regions indicate superconducting states (blue; SC1), superconducting-like resistance drops with finite resistance (light blue; SC2), the $\nu = 1$ Mott insulator (red; Mott1), an additional insulator on the electron side of $\nu = 1$ (yellow; Mott2), a weaker correlated insulating state near $\nu = 1/2$ (orange; Corr), generalised Wigner crystals (brown; GWC), van Hove singularity-dominated regimes (light grey; vHS1, light purple; vHS2), and the intervalley coherent antiferromagnet (dark grey; IVC-AFM). The blue area marked with diagonal lines is thought to be SC1, but the resistance does not reach zero and exhibits a noisy pattern. Upon warming, SC2 regions disappear and SC1 regions shrink (**c**), and at 1 K, most features have disappeared except the Mott and GWC states (**d**). The dashed line in **d** marks the locus of the calculated vHS.

**Phase diagram at $B = 0$**

We performed electrical transport measurements on a dual-gated $TTWSe_2$ device with a twist angle of 3.87°. **Fig. 1a** shows a schematic illustration of the dual-gated device structure. The twist angle was calibrated using quantum Hall oscillations (**Methods**, **Extended Data Fig. 2**). Ohmic contacts were achieved using a local contact gate (CG) to heavily hole-dope the device region in contact with Pt electrodes[14]. Top and bottom gate voltages were controlled simultaneously to tune the carrier density, or equivalently the moiré hole filling factor $\nu$, and vertical displacement field $D$ independently.

The measured longitudinal resistance $R(\nu, D)$ at 1 K (**Fig. 2d**) closely resembles the calculated DOS map, showing enhanced resistance at layer-hybridised regions and around the expected locations of vHS, reflecting the enhanced scattering rate possibly due to the large DOS. Furthermore, even at 1 K, correlated states not predicted by the single-particle picture are already emerging. Firstly, strong insulating states exist at $\nu = 1/3$ and 1/4 in layer-hybridised regions. We attribute these states to generalised Wigner crystals (GWC in **Fig. 2b**) favoured in the presence of long-range Coulomb interactions between neighbouring sites of the moiré lattice, which have been observed in other TMD moiré systems[15,16]. At half filling ($\nu = 1$), an insulating state (Mott1 in **Fig. 2b**) exists across a wide range of the displacement field. The strong-coupling description of this state is a Mott insulator on a triangular lattice[3,17]. Within triangular-lattice Hubbard/Heisenberg descriptions, a 120° antiferromagnetic Mott insulator is a natural candidate ground state, although competing spin-reorganised states have also been discussed[17–19]. In contrast to the cases in twisted bilayer $WSe_2$, where analogous insulators were observed only in a narrow range of displacement field near the intersections between half filling and the vHS[7] or the intervalley coherent antiferromagnetic order (IVC-AFM)[10], Mott1 remains robust throughout the explored $D$-field range including $D = 0$ and is well separated from the vHS (near $\nu = 0.75$ at $D = 0$). This suggests that $TTWSe_2$ system accesses a strong-correlation regime such that the insulating state is not driven by the onset of magnetic order. This is consistent with the picture that holes are more strongly confined by deep middle-layer moiré potential minima, thus reducing the effective hopping amplitude and bandwidth (**Fig. 1d, e** and **Extended Data Fig. 4**), stabilising the Mott insulating state.

As the temperature is lowered from **Fig. 2d** to **Fig. 2a**, features associated with interparticle interactions become clearer and additional correlated phases emerge. Detailed features observed at base temperature are summarised in **Fig. 2b.** The first notable feature is a broad region of a superconducting phase (SC1 in **Fig. 2b**) near Mott1. The robust superconductivity that appears upon slight doping of the half-filled correlated insulator resembles the phenomenology of high-$T_c$ superconductors[20,21] and contrasts with the case of $tWSe_2$, where superconductivity appears in proximity to the boundary of the intervalley coherent antiferromagnet order (IVC-AFM in **Fig. 2b**)[7,9,10]. At $0.93 < \nu < 0.96$ and $1.03 < \nu < 1.05$, SC1 shows substantial resistance fluctuations, suggestive of mesoscopic inhomogeneity or phase coexistence near the Mott transition in frustrated Hubbard models[22,23]. These fluctuations are reproducible to some degree in back-and-forth gate sweeps, but their pattern changes after a thermal cycle to room temperature (see **Extended Data Fig. 8**).

Between SC1 and Mott1, we also discover another relatively weak insulating state (Mott2 in **Fig. 2b**) and an insulator–insulator phase boundary. This state is characterised as a distinct insulator (see **Extended Data Fig. 9**), rather than a simple crossover before superconductivity appears upon further doping. Such intermediate insulating states between a metal and a 120° AFM Mott insulator have been repeatedly discussed in the half-filled triangular-lattice Hubbard model[24–29]. Candidates for such an intermediate insulating phase include a nonmagnetic Mott insulator[24,25], a gapless spin-liquid-like paramagnetic insulator[26], a chiral or gapped quantum spin liquid with emergent scalar-chiral order[27,28], or a more general spin-reorganised Mott state with chiral and stripy antiferromagnetic correlations competing with conventional 120° AFM order[29]. Distinguishing these possibilities and clarifying the relationship between the two insulating states and superconductivity present intriguing questions and call for further experimental and theoretical investigations.

In the $0.5 < \nu < 0.9$ region, several metallic states exhibit higher resistance than neighbouring regions. Firstly, regarding the high-resistance region observed at high *D*-field, we attribute this to the Fermi-surface-reconstructed metal induced by IVC-AFM (IVC-AFM in **Fig. 2b**), which was also observed in the twisted bilayer $WSe_2$ system[10]. Bare particle–hole susceptibility calculations show significant intervalley susceptibility near $\boldsymbol{q} = \kappa$ at high *D*-field (see **Fig. 4i**, **Extended Data Fig. 12**). Also, this region shrinks under a perpendicular magnetic field $B_z$ (**Extended Data Fig. 5a–c**). Such weakening under $B_z$ is naturally compatible with the IVC-AFM scenario, since, due to the spin–valley locking, the valley-Zeeman field tends to detune in-plane $\sigma_{x,y}$ intervalley coherence and favour $\sigma_z$ valley polarisation.

As *D*-field is lowered, vHS shifts away from the κ and κ′ points, and the intervalley nesting condition at κ can no longer be satisfied, causing the IVC-AFM to disappear. In this region, areas where the resistance is slightly higher than the surroundings appear near $\nu = 0.75$ and 0.5 (vHS1, vHS2 in **Fig. 2b**). vHS1 is close to the position of the vHS in the continuum model calculation, whereas the existence of the vHS2 region is not predicted by the calculation. Furthermore, along the quarter-filling side boundary of vHS2, regions where resistance is strongly suppressed appear centred around $D = \pm$ 120 mV nm$^{-1}$ (SC2 in **Fig. 2b**). As *D*-field approaches zero, SC2 regions disappear and an insulating state emerges at quarter filling (Corr in **Fig. 2b**). These states will be discussed in more detail later (see **Fig. 4** and the section therein).

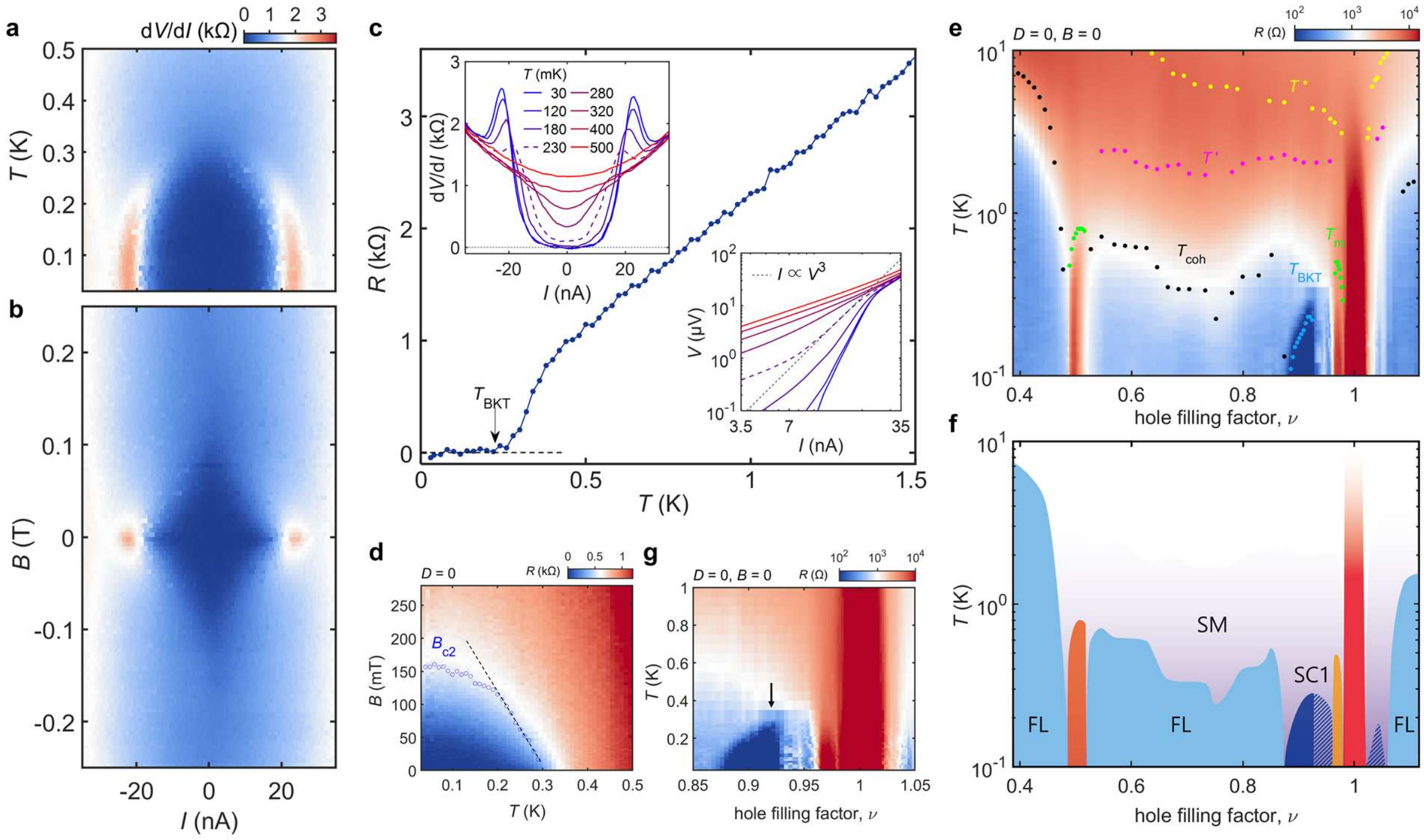

**Fig. 3 | Superconductivity near half moiré filling.**
**a, b**, differential resistance d$V$/d$I$ as a function of dc bias current $I$ and $T$ at $B$ = 0 (**a**), $I$ and $B$ at $T$ = 30 mK (**b**). **c,** Zero-bias resistance $R$ versus $T$ at $B$ = 0 and $D$ = 0. Upper-left inset, representative linecuts from **a**. Lower-right inset, corresponding $I$–$V$ curves. The curve at $T_{BKT} \approx 230$ mK is shown as a dashed line in both insets. **d,** Zero-bias resistance as a function of $T$ and $B$ at $D$ = 0. Blue circles mark the upper critical field $B_{c2}$, and the dashed line is a Ginzburg-Landau fit of $B_{c2}(T) = \frac{\Phi_0}{2\pi\xi^2(0)}\left(1 - \frac{T}{T_c}\right)$, yielding a coherence length of about 31 nm. **e,** Longitudinal resistance $R$ as a function of $\nu$ and $T$ at $D$ = 0 and $B$ = 0, defining the transport phase diagram. Black dots denote $T_{coh}$, marking the boundary of the Fermi-liquid regime. Magenta dots denote $T'$, marking the upper boundary of the strange-metal regime. Cyan dots denote the BKT transition temperature $T_{BKT}$ of SC1. Yellow dots denote the temperature $T^*$ at which the resistance, after a broad saturation, reaches a maximum and then decreases at higher temperature. Green dots denote the temperature $T_m$ at which the resistance of the insulating state reaches a minimum before increasing again. **f,** Schematic phase diagram corresponding to **e**. Cyan, Fermi liquid (FL); blue, superconducting state (SC1); purple, strange metal (SM); orange, yellow and red, distinct correlated insulating states identified in **Fig. 2b**. **g**, Zoomed-in image of **e** near $\nu$ = 1, showing a clear boundary between Mott2 and Mott1. The black arrow denotes $\nu$ = 0.92 where data in **a**–**d** were obtained.

**Superconductivity near the Mott insulator at $\nu = 1$**

To characterise the superconducting phase near half filling (SC1), we studied its response to bias current $I$, temperature $T$, and perpendicular magnetic field $B$ at $\nu = 0.92$ and $D = 0$. **Fig. 3a** shows the differential resistance $dV/dI$ as a function of $I$ and $T$ at $B = 0$. A finite critical current, approximately 20 nA at 30 mK, is required to destroy the zero-resistance state and it decreases monotonically as the temperature rises. We estimated the Berezinskii–Kosterlitz–Thouless (BKT) transition temperature as $T_{BKT} \approx 230$ mK, based on the temperature at which the nonlinear $I$–$V$ dependence becomes $V \propto I^3$ (insets of **Fig. 3c**). Similarly, **Fig. 3b** shows $dV/dI$ as a function of $I$ and $B$ at $T = 30$ mK and the critical current vanishes rapidly with increasing $B$. **Fig. 3c** shows the temperature dependence of zero-bias resistance at $B = 0$. The resistance decreases linearly upon cooling, then begins to drop rapidly around 300 mK, reaching the zero-resistance state near 200 mK. All of the qualitative behaviours are consistent with superconductivity. **Fig. 3d** shows the zero-bias resistance versus temperature and magnetic field. The upper critical field $B_{c2}$ is defined by the intersection of linear fits to the vortex- and normal-state resistance (see **Extended Data Fig. 7**), and decreases monotonically with increasing temperature. Fitting $B_{c2}(T)$ to the Ginzburg-Landau (GL) form $\frac{\Phi_0}{2\pi\xi^2(0)}\left(1-\frac{T}{T_c}\right)$ gives a superconducting coherence length of approximately 31 nm, about six times the moiré period $a_M$, signalling the formation of tightly bound superconducting pairs.

**Fig. 3e** and **3f** show the doping- and temperature-dependent resistance and the $T$-$\nu$ phase diagram constructed from the extracted temperature scales for the stability of various phases (see **Extended Data Fig. 6, 9**). At $0.88 < \nu < 0.96$, SC1 forms a dome structure adjacent to the correlated insulators at $\nu = 1$. On heavier electron doping ($0.12 < \delta = 1 - \nu < 0.5$), the resistance exhibits the $T^2$ dependence characteristic of the Fermi liquid. Above $T_{BKT}$ or $T_{\mathrm{coh}}$, the resistance exhibits a strange metal-like linear temperature dependence. Upon further warming above $T'$, the system crosses into a bad-metal regime, where the temperature dependence deviates from linearity, gradually saturates and reaches a maximum at $T^*$. Converting $R$ to the sheet resistance $R_\square = R\frac{W}{L}$ ($W$: channel width, $L$: distance between the longitudinal voltage probes) reveals that the maximum $R_\square$ at $T^*$ exceeds the Mott–Ioffe–Regel (MIR) limit $R_\square^{MIR} \approx \frac{2}{g}\frac{h}{e^2}\frac{1}{k_F l} \approx 25.8$ kΩ ($g = 2$: spin–valley degeneracy, $k_F$: Fermi wavevector, $l$: mean free path). This indicates that the mean free path becomes shorter than the moiré period (i.e. $k_F l \approx 1$) such that the coherent quasiparticle transport picture is no longer valid (see **Extended Data Fig. 6f**). These observations are consistent with the picture of a crossover from a correlated metal with coherent quasiparticles to an incoherent bad metal[30,31]. Beyond $T^*$, the resistance decreases. One plausible interpretation for this is an electronic Pomeranchuk effect: in contrast to the itinerant quasiparticle regime, local moments on triangular lattice can emerge in this incoherent regime and favour a more localised, entropy-rich Mott-like state upon heating[29,32,33]. If $T^*$ is taken as a proxy for the energy scale of the interaction-driven renormalised quasiparticle band[32,33], the relatively large ratio $T_{BKT}/T^* \sim 0.05$ is in line with the strong-pairing regime[7]. Taken together, these observations indicate that SC1 develops from a strongly correlated electronic background.

On the other hand, it should be noted that this effective Fermi-temperature scale $T^*$, a few Kelvin across the filling factors $0.5 < \nu < 1.1$, is significantly smaller than the single-particle Fermi temperature $T_F \gtrsim 150$ K estimated from the calculated bandwidth of ~ 25 meV (**Fig. 1d**). Therefore, using the single-particle $T_F$ for comparison may lead to a different conclusion about the pairing regime[10].

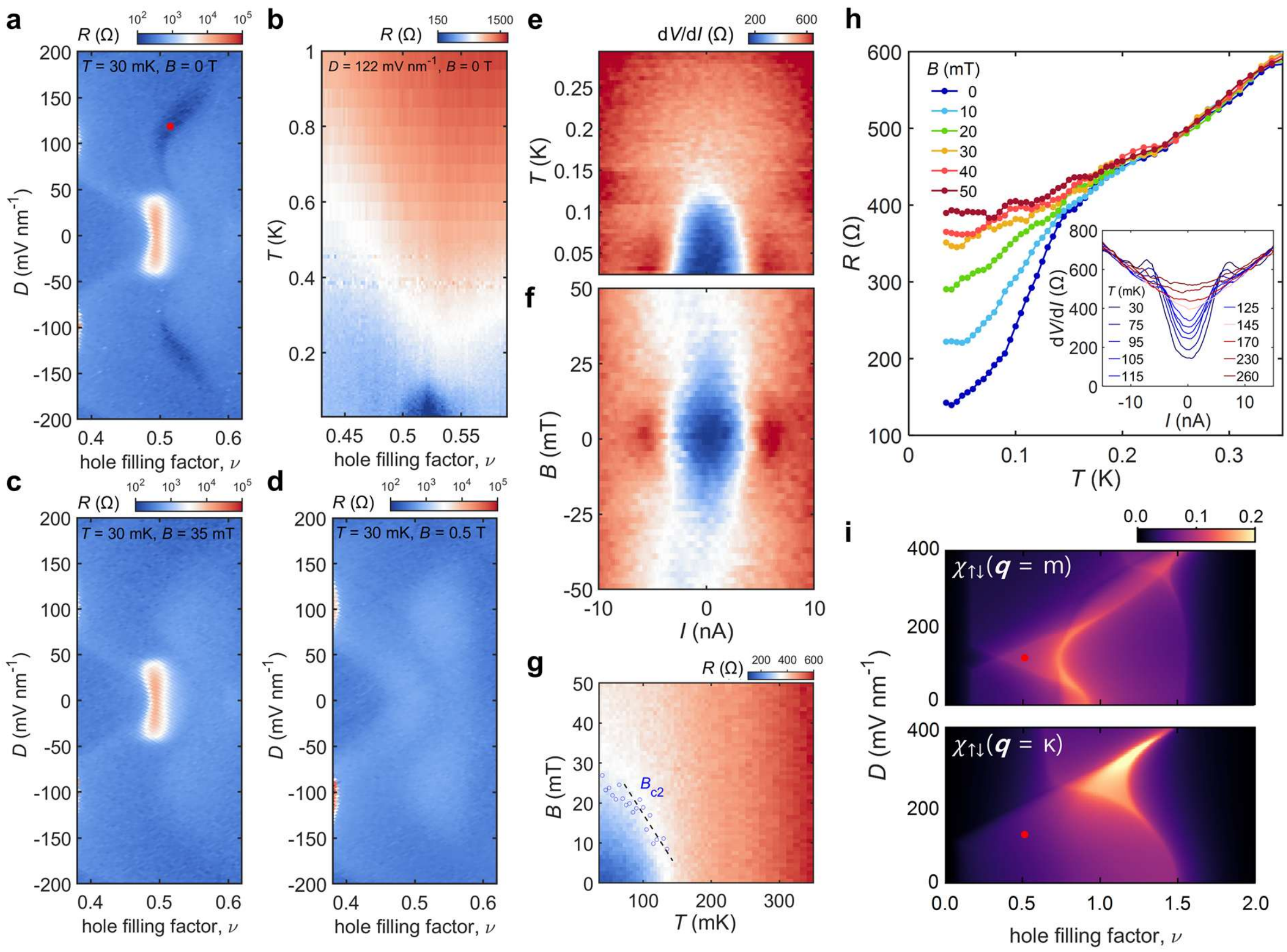

**Fig. 4 | Signatures of superconductivity and correlated phases near quarter filling.**
**a,** Zoomed-in view of **Fig. 2a** near $\nu = 1/2$. The red dot indicates the point at $\nu = 0.519$, $D = 122$ mV nm$^{-1}$ where data in **e**–**h** were obtained. **b,** Longitudinal resistance $R$ as a function of $\nu$ and $T$ at $B = 0$ and $D = 122$ mV nm$^{-1}$. SC2 forms a dome structure between metallic states. **c**,**d**, The same area as in **a** at $B = 35$ mT (**c**) and 0.5 T (**d**). As $B$ is increased, SC2 and the correlated insulator disappear. **e–h,** Characterisation of SC2 at $D = 122$ mV nm$^{-1}$. **e,f**, Differential resistance d$V$/d$I$ as a function of dc bias current $I$ and $T$ at $B = 0$ (**e**), $I$ and $B$ at $T = 30$ mK (**f**). **g**, zero-bias resistance as a function of $T$ and $B$. The blue circles represent the critical field estimated using the same method as in **Fig. 3d.** The dashed line is the GL fit yielding a coherence length of about 90 nm. **h,** Representative linecuts of **g**, showing progressive suppression of the sharp resistance drop with increasing $B$. Inset, d$V$/d$I$ versus $I$ at selected temperatures from **e**. **i**, Calculated intervalley particle–hole susceptibility as a function of $\nu$ and $D$ at momentum transfer $\boldsymbol{q}$ = m (upper), $\boldsymbol{q}$ = κ (lower). Red dots mark the same point in **a**.

**Correlated insulator near $\nu = 1/2$ and signatures of a superconducting phase**

At quarter filling, the region near $D = 0$ coincides with the Lifshitz transition boundary between the layer-polarised and layer-hybridised regions; as shown **in Fig. 4a**, an insulating state emerges at this point (Corr in **Fig. 2b**). Considering the commensurability about the triangular lattice, we attribute this phase to a stripe-ordered phase with charges occupying half of the moiré lattice sites, which has been previously reported in TMD heterobilayers[34,35]. However, unlike the GWC states observed at $\nu = 1/3$ and 1/4, this insulating state is highly fragile: it begins to disappear around 0.5 K (**Fig. 3e** and **Extended Data Fig. 6b**), and is suppressed by a small magnetic field (**Fig. 4d**, **Extended Data Fig. 5**). These characteristics suggest that it is unlikely to be a purely charge-driven crystal; instead, magnetic order may play a significant role in stabilising the phase.

Upon increasing the magnitude of displacement field, a pair of suppressed-resistance regions (SC2) emerges. Rather than being tied to quarter-filling like the insulator at $D = 0$, SC2 appears at the boundary of the interaction-driven metallic state (vHS2) formed adjacent to quarter filling (**Fig. 4a, b**). Although SC2 fails to reach zero resistance, its disappearance under weak magnetic fields in **Fig. 4c**, together with the nonlinear transport characteristics analysed in **Fig. 4e–f** and the temperature and magnetic field responses of the zero-bias resistance in **Fig. 4g–h**, strongly suggest a superconducting origin. Possible reasons for the residual resistance include the base temperature remaining above the BKT transition temperature, or finite-size effects and spatial inhomogeneity within the device[36]. Similar non-zero-resistance superconducting signatures, often appearing near symmetry-broken correlated metallic states, have also been reported in rhombohedral graphene systems[37–39].

Since the vHS2 region is not predicted from single-particle calculations, it can be considered a type of Fermi-surface-reconstructed metal arising from magnetic ordering. This raises the possibility that spin fluctuations near the boundary of this magnetically ordered phase contribute to the pairing mechanism[40–42] of SC2. While such magnetic order is a commensurate $\boldsymbol{q} = \kappa$ IVC-AFM in $tWSe_2$[10], it is unlikely that vHS2 corresponds to $\boldsymbol{q} = \kappa$ IVC-AFM for the case of SC2. This is because, at $|D| < 200$ mV nm$^{-1}$, the vHS is displaced from the κ point, suppressing the intervalley susceptibility $\chi_{\uparrow\downarrow}(\boldsymbol{q} = \kappa)$, while $\chi_{\uparrow\downarrow}(\boldsymbol{q} = \mathrm{m})$ remains relatively large (**Fig. 4i**, **Extended Data Fig. 12**).

Near $D = \pm 120$ mV nm$^{-1}$, where SC2 is most prominent, the energies at the κ and κ′ points are balanced, placing the vHS at the m point and making the band structure similar to that of $tWSe_2$ at $D = 0$ (**Extended Data Fig. 3, 12**). As the $D$ field deviates from $\pm 120$ mV nm$^{-1}$ near quarter filling, SC2 regions shrink and $\chi_{\uparrow\downarrow}(\boldsymbol{q} = \mathrm{m})$ decreases. Theoretical calculations for $tWSe_2$ predict that the valley-polarised ferromagnetic (VP-FM) order, $\boldsymbol{q} = \mathrm{m}$ striped IVC-AFM order, and incommensurate IVC-AFM orders in between can develop near the vHS in such a low-$D$-field region[40]. Among these orders, vHS2 is also unlikely to be VP-FM, because, although the longitudinal magnetic susceptibility $\chi_{zz}(\boldsymbol{q} = 0) \propto \frac{h}{e}\frac{\partial(n_\uparrow - n_\downarrow)}{\partial B} \approx \frac{E_Z}{E_C} = \frac{g^* m^*}{m_0}$ is highly enhanced relative to the monolayer[43] (see **Methods** and **Extended Data Fig. 5i**), no hysteresis was observed at vHS2 under a vertical magnetic field and the vHS2 feature weakens above $B \approx 1$ T (**Extended Data Fig. 5**).

Taking these observations together, we conjecture that $\boldsymbol{q} = \mathrm{m}$ striped IVC-AFM spin fluctuations may contribute to the pairing mechanism of SC2. It is noteworthy that, since the vHS2 region is located near quarter filling, stripe order is likely to be favoured from the perspective of triangular lattice commensurability. However, while bare susceptibility calculations can suggest possible candidates for magnetic order, this alone is insufficient to predict which order will actually develop. While $\boldsymbol{q} = \kappa$ IVC-AFM seems to appear near the vHS in the high $D$-field region, resembling $tWSe_2$ (**Fig. 2b**), the vHS2 region ($\nu \sim 0.5$) is well separated from the single-particle vHS ($\nu \sim 0.75$), and no corresponding feature has been observed in any reported transport measurements of $tWSe_2$ at angles ranging from 2.1° to 5.0°[7–10]. Therefore, further theoretical calculations, such as functional renormalisation group (FRG) with Hartree-Fock calculations, are required to investigate which order can develop in the low $D$-field region in $TTWSe_2$, how band reconstruction occurs to give rise to features such as vHS2 near quarter filling, and what the nature of the superconductivity that arises alongside it is.

**Discussion and Conclusion**

An important issue requiring further theoretical and experimental investigation is the topology of the relevant topmost moiré band, since it determines whether the low-energy effective model is a single-orbital Hubbard model or a multi-orbital Kane-Mele-Hubbard model. Many theoretical studies of $tWSe_2$ have predicted that the topmost band is topological[4,44,45], and predictions regarding magnetic ordering and superconductivity are also based on a multi-orbital Wannier description of the system[40,46,47]. However, band topology calculations are highly sensitive to the choice of continuum model parameters[17]. We adopted a set of continuum model parameters extracted from an STM experiment[48] because it best matched the experimental ($\nu$, $D$) phase diagram with the calculated DOS map (**Extended Data Fig. 11**). For this set of parameters, the topmost band was calculated to be trivial (**Extended Data Fig. 4**). The robust Mott insulating state at half-filling and the absence of hysteresis or signatures of the quantum anomalous Hall effect are consistent with the single-band Hubbard model description of the system. However, this alone does not completely rule out the possibility that the topmost valence band possesses a non-zero Chern number.

In conclusion, we observed transport signatures of multiple superconducting phases in the $TTWSe_2$ system, which are expected to be distinct from the superconductivity in $tWSe_2$, together with a rich phase diagram including correlated insulators and correlated metals signalling a strongly correlated regime. Our results suggest that the $TTWSe_2$ system is a compelling platform for studying Mott physics, superconductivity, and strongly correlated states in the triangular-lattice Hubbard model with high tunability.

## Methods

### Device fabrication

**Extended Data Fig. 1** shows an optical image of the device. The heterostructure was assembled using a standard dry-transfer technique. First, hBN and graphite flakes were sequentially picked up using a polycarbonate (PC) film mounted on a polydimethylsiloxane (PDMS) stamp and released onto a Si/$SiO_2$ substrate to form a prestack. Ti/Pt contacts (2 nm/5 nm) were then deposited, followed by mechanical cleaning using an atomic force microscope (AFM). A monolayer $WSe_2$ flake was subsequently cut mechanically using an AFM tip[49]. The trilayer stack was assembled by sequentially picking up hBN, a $WSe_2$ monolayer, a second $WSe_2$ monolayer rotated by the target angle, and a third $WSe_2$ monolayer rotated back to 0°, followed by release onto the prestack. After an additional AFM cleaning step, Ti/Pt (2 nm/5 nm) was deposited locally near the Pt-TT$WSe_2$ contact region to gate the contact area. The sample was then cleaned again by AFM, and a top stack consisting of hBN, graphite, and hBN was transferred onto the heterostructure. Finally, Ti/Au electrodes (5 nm/45 nm) were deposited to contact the Pt leads as well as the top and bottom graphite gates.

### Electrical transport measurements

Electrical transport measurements were carried out in a dilution refrigerator equipped with a 14 T superconducting magnet (Bluefors LD250). To reduce the electron temperature, RC/RF filters were installed at the mixing chamber, and a sapphire heat sink was mounted on the sample probe. A negative voltage about −5 V was applied to the contact gates (CG) to activate the contacts. Resistance measurements were performed using a standard low-frequency lock-in technique at 5.555 Hz. A 100 MΩ resistor was connected in series with the voltage source to approximate a 1 nA (RMS) AC current source from a 100 mV (RMS) excitation. The source-drain current and the voltage drop across the voltage probes were recorded simultaneously.

### Twist-angle determination

First, the longitudinal resistance was measured by sweeping the top and bottom gates at the base temperature. The ratio between $c_t$ and $c_b$ was then determined from the equations below.

$$n = \frac{c_t V_t + c_b V_b}{e} = \frac{c}{e}\{(\cos\alpha)V_t + (\sin\alpha)V_b\} \equiv \frac{c}{e}V_x$$

$$D = \frac{-c_t V_t + c_b V_b}{2\epsilon_0} = \frac{c}{2\epsilon_0}\{-(\cos\alpha)V_t + (\sin\alpha)V_b\} \equiv \frac{c}{2\epsilon_0}V_y$$

$$c = \sqrt{{c_t}^2 + {c_b}^2}, \qquad \alpha = \operatorname{atan}\left(\frac{c_b}{c_t}\right)$$

The longitudinal resistance was then measured as a function of magnetic field $B$ and $V_x$. In the quantum Hall regime, the capacitance per unit area $c$ is determined from the value of $V_x$ at which each Landau level is filled. We then verified whether this value is consistent with that predicted based on the thickness of the hBN. Subsequently, the moiré density was calibrated from the carrier density at which features characteristic of commensurate moiré filling appear (see **Extended Data Fig. 2**).

**Continuum model calculation**

The low-energy hole bands of alternating twisted trilayer $WSe_2$ were modelled within a continuum k·p framework obtained by extending the standard twisted-bilayer TMD description[13] to trilayers[11]. Owing to spin–valley locking near the valence-band maximum of monolayer $WSe_2$, the K and K′ sectors were treated independently and related by time-reversal symmetry. For a given valley, the Hamiltonian was written in the three-layer basis and included parabolic monolayer dispersions within an effective-mass approximation. Layer-dependent intralayer moiré potentials and nearest-neighbour interlayer moiré tunnelling were retained to lowest harmonic order, while direct tunnelling between the two outer layers was neglected. To represent the alternating geometry, the valley pockets of the two outer layers were taken at the same moiré-Brillouin-zone corner, whereas that of the middle layer was taken at the opposite corner. A perpendicular electric field was incorporated as opposite electrostatic shifts on the two outer layers, with the middle layer left unchanged.

$$H_{\uparrow} = \begin{pmatrix} -\frac{\hbar^2(\boldsymbol{k}-\boldsymbol{\kappa})^2}{2m^*} + \Delta_+^{12} + \frac{V_z}{2} & \Delta_T^{12}(\boldsymbol{r}) & \\ \Delta_T^{12,\dagger}(\boldsymbol{r}) & -\frac{\hbar^2(\boldsymbol{k}-\boldsymbol{\kappa}')^2}{2m^*} + \Delta_-^{12} + \Delta_+^{23} & \Delta_T^{23}(\boldsymbol{r}) \\ & \Delta_T^{23,\dagger}(\boldsymbol{r}) & -\frac{\hbar^2(\boldsymbol{k}-\boldsymbol{\kappa})^2}{2m^*} + \Delta_-^{23} - \frac{V_z}{2} \end{pmatrix}$$

Here, **k** and **r** denote the hole momentum and position, respectively, and $\boldsymbol{\kappa}$ and $\boldsymbol{\kappa'}$ are the corners of the moiré Brillouin zone. The effective mass of the topmost valence band was taken to be $m^* \approx 0.45\ m_0$, where $m_0$ is the free-electron mass[50]. In the lowest harmonic approximation, the intralayer moiré potential and interlayer tunnelling are described as follows.

$$\Delta_\pm^{ll'}(\boldsymbol{r}) = 2V \sum_{j=1,3,5} \cos\left(\boldsymbol{G}_j^{ll'} \cdot \boldsymbol{r} \pm \psi\right)$$

$$\Delta_T^{ll'}(\boldsymbol{r}) = w\left(1 + e^{-i\boldsymbol{G}_2^{ll'}\cdot \boldsymbol{r}} + e^{-i\boldsymbol{G}_3^{ll'}\cdot \boldsymbol{r}}\right)$$

The reciprocal lattice vectors of the moiré pattern at the 1−2 and 2−3 interfaces are defined by $\boldsymbol{G}_j^{ll'} = \boldsymbol{b}_j^{l'} - \boldsymbol{b}_j^{l}$ where $\boldsymbol{b}_j^{l}$ is the *j*th reciprocal lattice vector of the *l*th monolayer $WSe_2$. $G_j^{ll'}$ is obtained by rotating $G_1^{ll'} = \left(\pm\frac{4\pi}{\sqrt{3}a_M}, 0\right)$ (+ for (*l, l'*) = (1, 2), - for (*l, l'*) = (2, 3)) anticlockwise by an angle $\frac{(j-1)}{3}\pi$. The interlayer potential difference $V_z$ was converted to the $D$ field using the relation $D = \frac{V_z}{\left(\frac{\epsilon_{hBN}}{\epsilon_{TMD}} ed\right)}$. The dipole moment $\frac{\epsilon_{hBN}}{\epsilon_{TMD}} ed \approx 0.26\ e$ nm was adopted from the experimental value obtained from optical spectroscopy on the moiré exciton[32].

The band structure was obtained by representing the real-space operators in a plane-wave basis and numerically diagonalising the Hamiltonian. DOS was calculated from the resulting band dispersion, and the layer polarisations were extracted from the squared norm of layer-resolved components of the eigenvectors. $\boldsymbol{\psi} = (\psi_T, \psi_M, \psi_B)$, $p_l = |\psi_l|^2/|\boldsymbol{\psi}|^2$ ($l$ = T, M, B). For presentation, layer polarisation was normalised by its maximum component. $p_{l,\mathrm{normalised}}(k) = p_l(k)/\max\{p_T(k), p_M(k), p_B(k)\}$

Importantly, because systematic studies of twisted trilayer $WSe_2$ remain limited, we tested several continuum model parameter sets for twisted bilayer $WSe_2$ from the literature[4,11,46,48] and selected the parameter set (V,ψ,w) = (13.6 meV, 49.1°, 10.0 meV) estimated from an STM experiment[48] that best matched the boundary between the layer-polarised region and the layer-hybridised region on the DOS ($\nu$,D) map (see **Extended Data Fig. 11**).

To account for Zeeman splitting, we introduced a constant effective hole g-factor, $g^*$, and applied opposite energy shifts to the zero-field K- and K′-valley bands obtained from the above Hamiltonian, such that the total valley splitting is $E_Z = 2g^*\mu_B B$, where $\mu_B$ denotes the Bohr magneton. We estimated $g^*$ from experimental monolayer values of $g^*m^*/m_0$ versus carrier density[43], taking $m^* = 0.45\ m_0$ and interpolating to the carrier density at the van Hove singularity of $TTWSe_2$, giving $g^* \approx 12.22$. The corresponding total Zeeman splitting per Tesla was estimated to be 1.415 meV $T^{-1}$. Although the use of a constant effective g-factor is a simplified treatment, it captures the essential features of the Fermi-surface structure (see **Extended Data Fig. 10**).

**Berry curvature and Chern number calculation**

Berry curvatures and Chern numbers were calculated from the cell-periodic Bloch eigenstates $|u_n(\boldsymbol{k})\rangle$ of each isolated moiré band using the gauge-invariant lattice formulation of Fukui, Hatsugai, and Suzuki[51]. On a uniform periodic $k$-space mesh, the normalised link variables are defined as $U_\mu(\boldsymbol{k}_{ij}) = \frac{\langle u_n(\boldsymbol{k}_{ij})|u_n(\boldsymbol{k}_{ij}+\hat{\mu})\rangle}{|\langle u_n(\boldsymbol{k}_{ij})|u_n(\boldsymbol{k}_{ij}+\hat{\mu})\rangle|}$ for $\mu$ = 1,2, and the Berry flux through each plaquette as

$$F_{12}(k_{ij}) = \arg\left[U_1(\boldsymbol{k}_{ij})U_2(\boldsymbol{k}_{ij}+\hat{1})U_1(\boldsymbol{k}_{ij}+\hat{2})^{-1}U_2(\boldsymbol{k}_{ij})^{-1}\right]$$

The Berry curvature was then approximated as $\Omega_n(\boldsymbol{k}_{ij}) = \frac{F_{12}(k_{ij})}{\Delta S_k}$, where $\Delta S_k$ is the area of one plaquette in the moiré Brillouin zone, and the Chern number was obtained from $C_n = (2\pi)^{-1}\sum_{ij} F_{12}(\boldsymbol{k}_{ij})$

**Wannier function construction and on-site interaction estimation**

From the isolated topmost moiré valence band, we constructed a localised single-band Wannier function by diagonalising the continuum Hamiltonian on a uniform $N_k \times N_k$ mesh of the moiré Brillouin zone, fixing a smooth gauge using a two-dimensional parallel-transport procedure. The layer-resolved real-space orbital was reconstructed as

$$W_{l,\boldsymbol{R}}(\boldsymbol{r}) = \frac{1}{\sqrt{N_k^2}}\sum_k e^{-i\boldsymbol{k}\cdot\boldsymbol{R}}\psi_{lk}(\boldsymbol{r}), \qquad \psi_{l\boldsymbol{k}}(\boldsymbol{r}) = \sum_{\boldsymbol{G}} u_{lk}(\boldsymbol{G})e^{i(\boldsymbol{k}+\boldsymbol{G})\cdot\boldsymbol{r}}$$

Here, $l$ = 1, 2, 3 labels the trilayer degrees of freedom, and the orbital was normalised such that $\sum_l \int d^2r\left|W_{l,\boldsymbol{R}}(\boldsymbol{r})\right|^2 = 1$.

As a rough estimate of the on-site interaction $U$, we calculated the screened Coulomb interaction by introducing the Rytova–Keldysh potential[52]. $\rho_l(\boldsymbol{r}) = \left|W_{l,\boldsymbol{0}}(\boldsymbol{r})\right|^2$ ($l$ = 1, 2, 3) is the layer-resolved density and $\rho_l(\boldsymbol{q})$ is the Fourier transform of $\rho_l(\boldsymbol{r})$. The on-site $U$ was calculated as follows.

$$U = \sum_{l,l'} \int \frac{d^2q}{(2\pi)^2} V(\boldsymbol{q}) e^{-q|z_l - z_{l'}|} \rho_l(\boldsymbol{q}) \rho_{l'}(-\boldsymbol{q}), \qquad V(q) = \frac{2\pi e^2}{q(\kappa + r_0 q)}$$

Here $z_l$ denotes the out-of-plane position of layer $l$, so that the factor $e^{-q|z_l - z_{l'}|}$ accounts for the finite separation between layers. For the dielectric environment used in the calculation, we adopted $\kappa = 4.5$ and $r_0 = 4.5$ nm that are commonly used in the $WSe_2$ magneto-exciton theory and experiments[53,54].

**Noninteracting static particle–hole susceptibility calculation**

To analyse the weak-coupling instabilities of the $TTWSe_2$ system, the noninteracting static particle–hole susceptibilities were calculated as follows.

$$\chi_{\sigma\tau}(\boldsymbol{q}) = \frac{1}{N} \sum_{k} \frac{f_\sigma(\boldsymbol{k}) - f_\tau(\boldsymbol{k} + \boldsymbol{q})}{E_\sigma(\boldsymbol{k}) - E_\tau(\boldsymbol{k} + \boldsymbol{q}) + i\eta}$$

where $E_\sigma(\boldsymbol{k})$ is the continuum model energy of the topmost band of σ ($= K_\uparrow, K'_\downarrow$) valley, $f_\sigma(\boldsymbol{k}) = \left\{1 + \exp\left(\frac{E_\sigma(\boldsymbol{k}) - \mu}{k_B T}\right)\right\}^{-1}$ is the Fermi occupation, and $\eta$ is a smearing parameter that regularises possible divergences and adjusts the contribution of states distant from the Fermi level. For **Extended Data Fig. 12**, $k_B T = 0.003$ meV, $\eta = 0.3$ meV were used.

**Data availability**

The data that support the findings of this study are available from the corresponding author upon reasonable request.

**Acknowledgements**
We thank Jeong Min Park for helpful discussions on the transport signatures of superconductivity near quarter filling and Moon Jip Park for discussions on theoretical models. This work was supported by the National Research Foundation of Korea grants funded by the Ministry of Science and ICT (Grant Nos. RS-2025-23525425, RS-2020-NR049536, and RS-2023-00258359), the Institute for Basic Science of Korea (Grant No. IBS-R009-D1), SNU Core Center for Physical Property Measurements at Extreme Physical Conditions (Grant No. 2021R1A6C101B418), and Creative-Pioneering Researcher Program through Seoul National University. K. W. and T. T. acknowledge support from the JSPS KAKENHI (Grant Nos. 21H05233 and 23H02052) and World Premier International Research Center Initiative (WPI), MEXT, Japan.

**Author contributions**
Y.J. and J.J. conceived the project. Y.J. fabricated samples, conducted measurements and performed theoretical calculations. K.W. and T.T. provided hBN crystals. Y.J. and J.J. analysed data and wrote the manuscript. J.J. supervised the overall project with input from all authors.

**Competing interests**
The authors declare that they have no competing interests.

## Extended Data figures

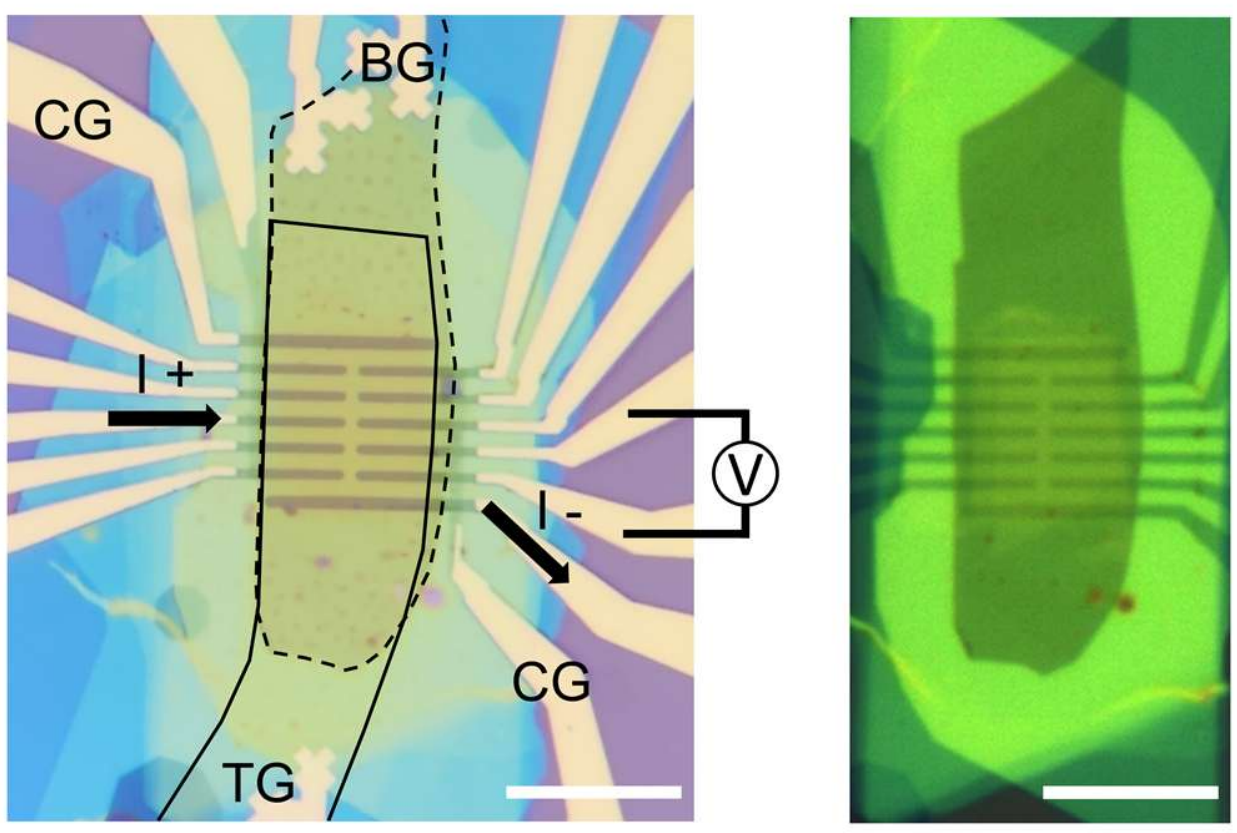

**Extended Data Fig. 1 | Optical image of the $TTWSe_2$ device and measurement configuration.**
Optical microscope image of a 3.87° $TTWSe_2$ device. The $TTWSe_2$ is situated in the region between the top gate (TG, graphite, solid black line) and the bottom gate (BG, graphite, dotted black line). To achieve an ohmic contact, contact-gate (CG) electrodes were additionally positioned between the TG and the $TTWSe_2$, overlapping the electrode. Most of the measurements described in the text were performed by applying a current from the I+ electrode to the I− electrode, as indicated in the figure, and measuring the voltage between the two electrodes marked with the voltmeter symbol. The device width is about 12 μm and the channel length is 2.4 μm. The scale bar is 15 μm.

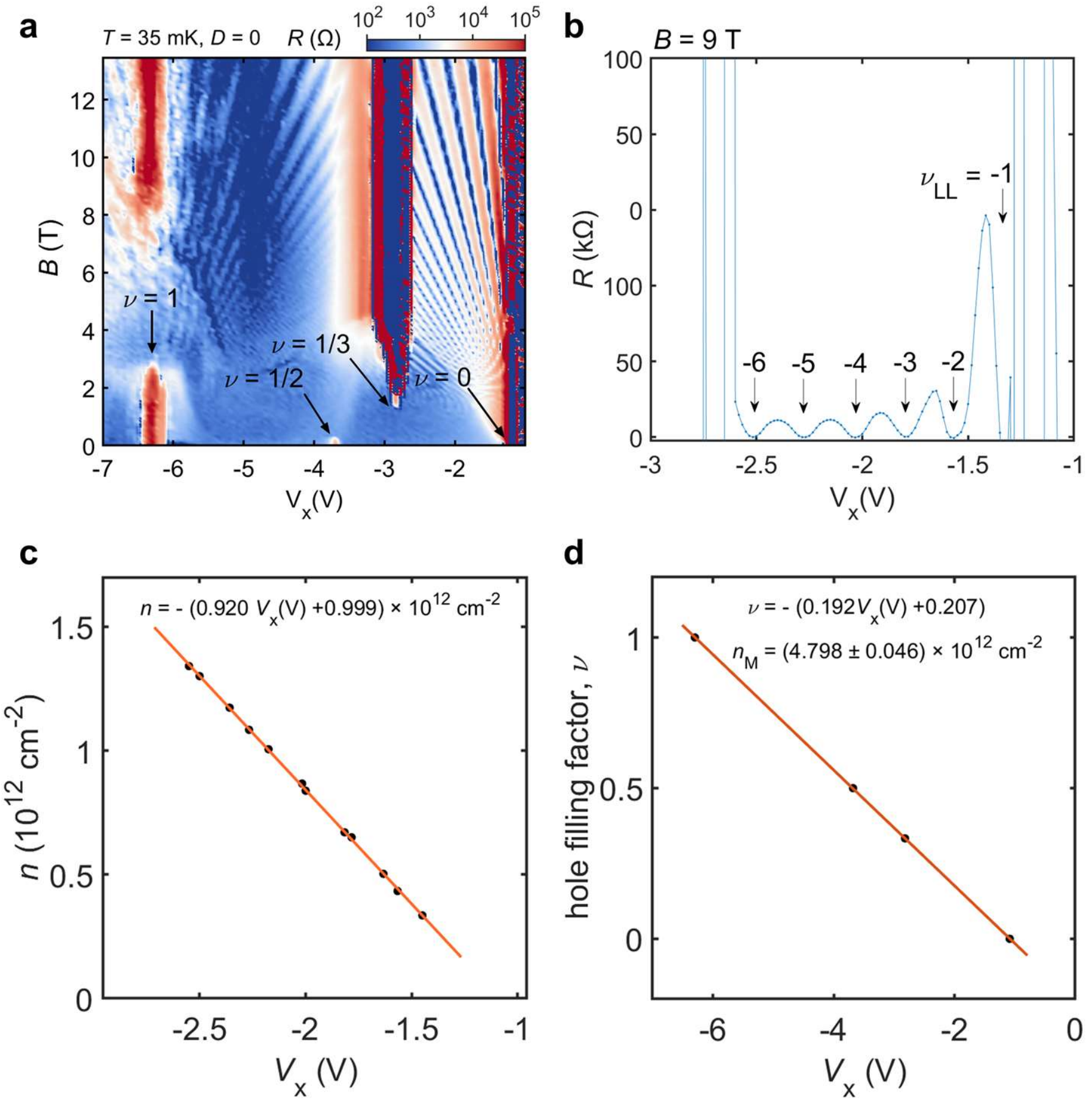

**Extended Data Fig. 2 | Determination of the twist angle.**

The twist angle was estimated using features observed in the Landau fan and commensurate moiré fillings. **a**, Longitudinal resistance $R$ as a function of $V_x$ and magnetic field at $T = 35$ mK and $D = 0$. **b**, A linecut at 9 T from **a**. For each Landau filling factor, the longitudinal resistance reaches a value close to zero at regular intervals. **c**, $V_x$ versus carrier density. By tracking the $V_x$ values at which the resistance reaches a local minimum at various magnetic fields, as shown in **b**, the carrier density as a function of $V_x$ was determined. **d**, $V_x$ versus moiré hole filling factor $\nu$. We estimated the $V_x$ values for each filling factor from the positions of features observed at commensurate fillings (1/3, 1/2, 1) at low magnetic fields, and compared these with the trend line calculated in **c** to determine the moiré density as $(4.798 \pm 0.046) \times 10^{12}$ cm$^{-2}$. By comparing this with the moiré density as a function of the twist angle in twisted $WSe_2$, the twist angle was estimated to be approximately 3.87°.

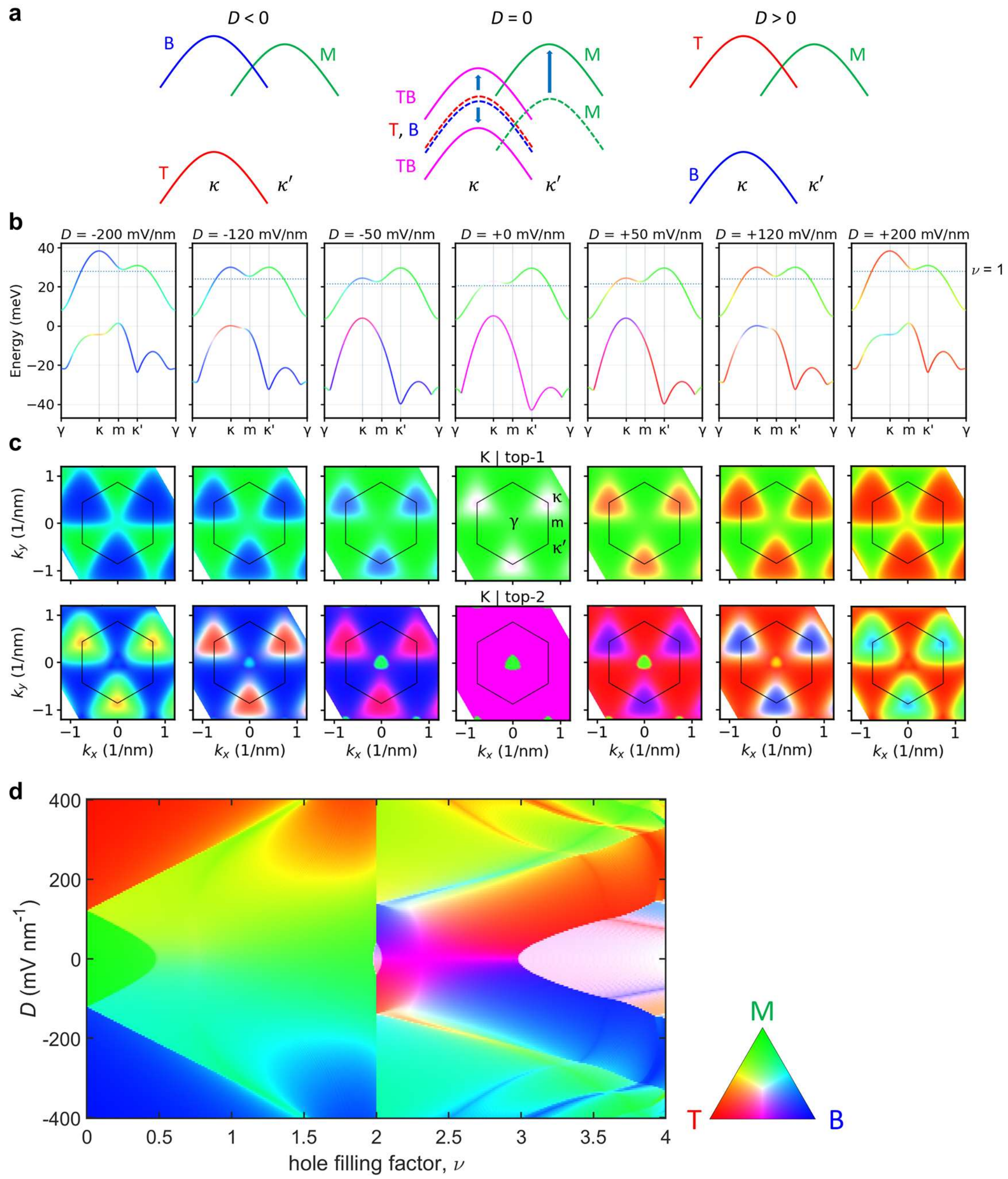

**Extended Data Fig. 3 | Continuum model calculation of the band structure of 3.9° TTWSe$_2$.**
**a**, Schematic diagram of the TTWSe$_2$ band structure described in the main text. In the absence of interlayer tunnelling and moiré potentials, the $K_\uparrow$-valley states of each monolayer are located at the moiré Brillouin zone corners with the same energy, as indicated by the dashed curves in the central diagram where $D = 0$. **b**, K-valley band structure at various vertical displacement fields. Near $D = 120$ mV nm$^{-1}$, the energies of the κ-valley and κ′-valley become similar, resembling the band structure of a twisted bilayer WSe$_2$ at $D = 0$. **c**, 2D map of layer polarisation in the topmost and secondary valence bands. For the κ-valley at $D = 0$, the schematic in **a** suggests that the top and bottom layers hybridise to form a magenta colour; however, the calculations show a colour closer to white due to the presence of the middle layer component. **d**, Layer polarisation as a function of the hole filling factor $\nu$ and the vertical displacement field $D$. Each point ($\nu$, $D$) represents the average of the layer polarisation of all states corresponding to that point.

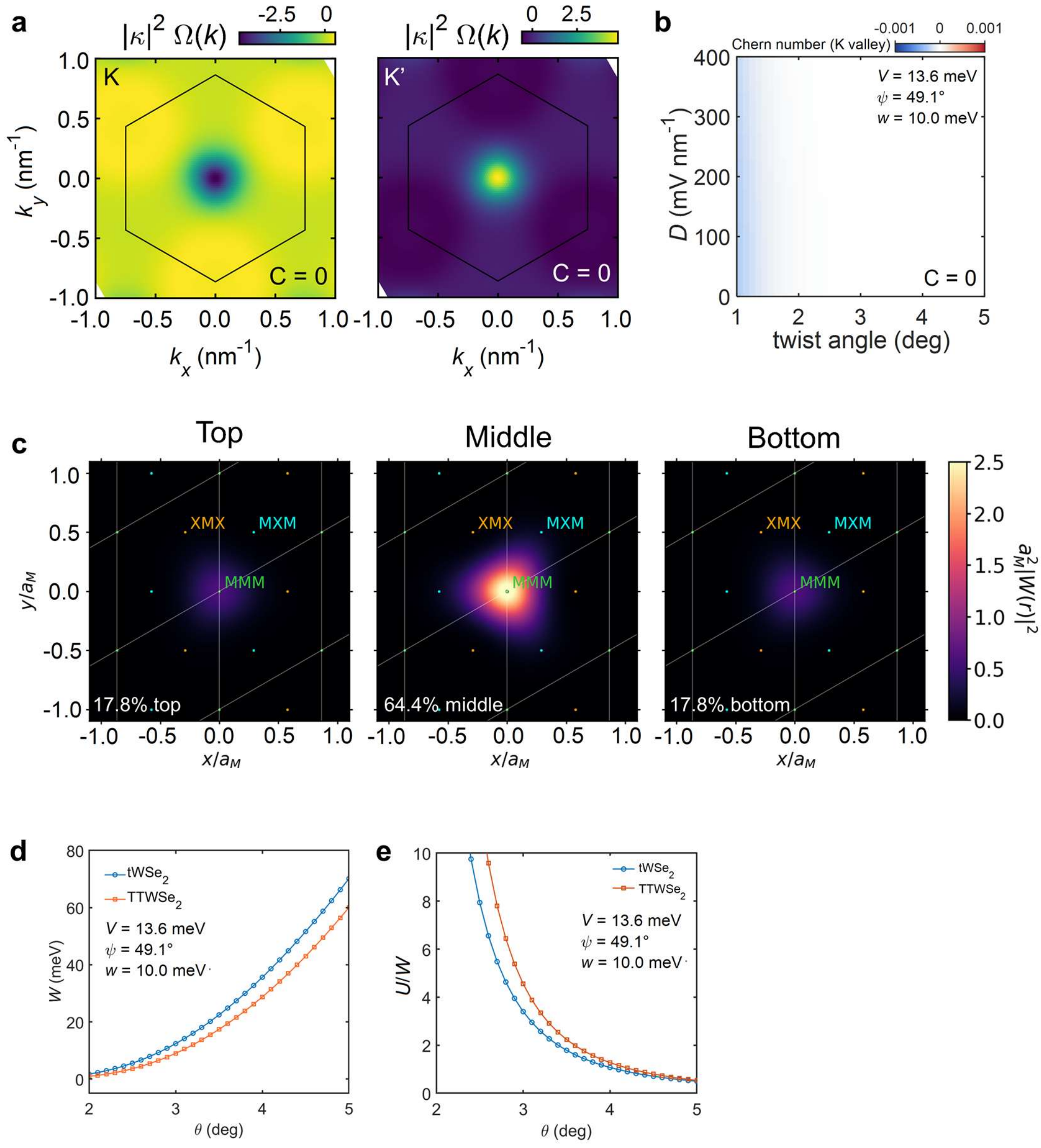

**Extended Data Fig. 4 | Berry curvature, Chern number and Wannier function of the topmost valence band of $TTWSe_2$.**

**a**, Continuum-model calculation of the scaled Berry curvature $|\kappa|^2\Omega(k)$ of the topmost valence band for $TTWSe_2$ at 3.9° and $D = 0$ (left, K-valley; right, K′-valley). The Berry curvature peaks near the γ point, and the Chern number—obtained by integrating the Berry curvature over the moiré Brillouin zone—is 0, indicating a trivial topology. **b**, The Chern number of the K-valley band as a function of the twist angle and the displacement field $D$. Under the continuum model parameter set used, the Chern number is always zero for the scanned ranges of twist angle and $D$, within numerical error. **c**, Wannier functions constructed from the topmost band at $D = 0$. The parallelogram represents the moiré unit cell, and the coloured points denote the high-symmetry sites MMM (green), XMX (orange) and MXM (cyan). In the trivial regime (C = 0), a single-orbital Wannier state localised at the MMM site can be constructed. The RMS spread $a_W$ was estimated to be 1.64 nm. **d,e**, Bandwidth $W$ (**d**) and $U/W$ (**e**) as a function of the twist angle $\theta$ for twisted bilayer $WSe_2$ ($tWSe_2$) and alternating twisted trilayer $WSe_2$ ($TTWSe_2$) at $D = 0$. The same continuum model parameter values were used in the calculations. $U$ was calculated using the Wannier function in **c** (**Methods**); a similar method was used for $tWSe_2$.

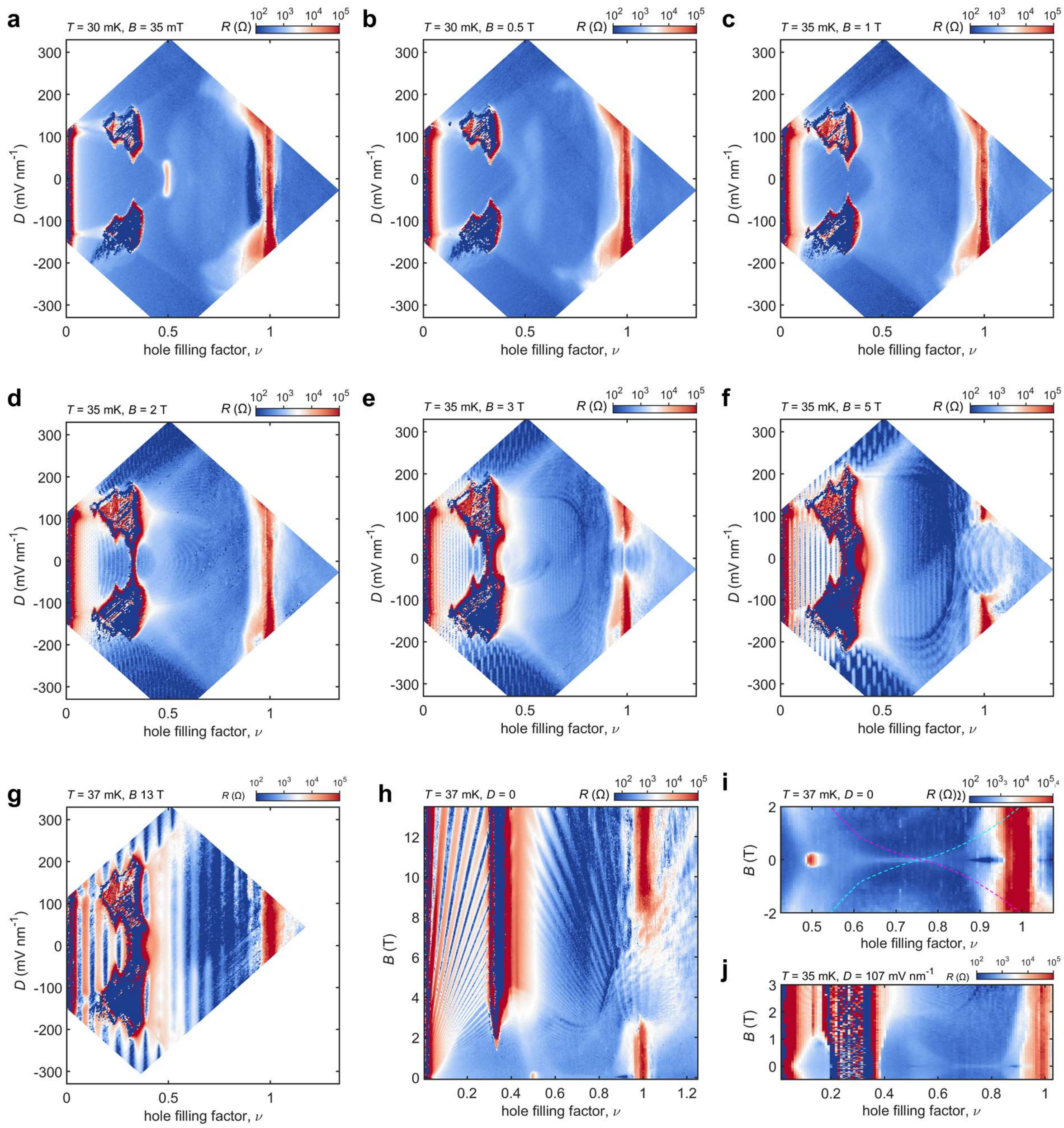

**Extended Data Fig. 5 | Magnetic-field responses at base temperature**

**a**–**g,** Longitudinal resistance as a function of $v$ and $D$ at various $B$. **a**, When a weak magnetic field of 35 mT is applied, the region of SC1 shrinks and SC2 disappears, resulting in a pattern very similar to that shown in **Fig. 2c**. However, in **a**, despite the small magnetic field, the vHS shifts significantly due to the very large spin–valley susceptibility, causing the peak near $v = 0.75$ to disappear. **b**–**c**, As the magnetic field is increased to 1 T, SC1 also disappears, and it can be seen that vHS shifts and IVC-AFM shrinks, leaving the region around $v = 0.75$ depleted. **d**–**g**, From a magnetic field of 2 T onwards, various patterns of quantum oscillations appear due to the Fermi surface structure. **h**–**j**, Longitudinal resistance as a function of $v$ and $B$ at $T = 37$ mK, $D = 0$ (**h**, **i**), and $T = 35$ mK, $D = 107$ mV nm$^{-1}$ (**j**). The dashed lines in **i** represent the position of single-particle vHS of K (magenta), K′ (cyan) valley with an effective g-factor $g^* = 12.22$. For detailed analysis of the Fermi surface structure, see **Extended Data Fig. 10.**

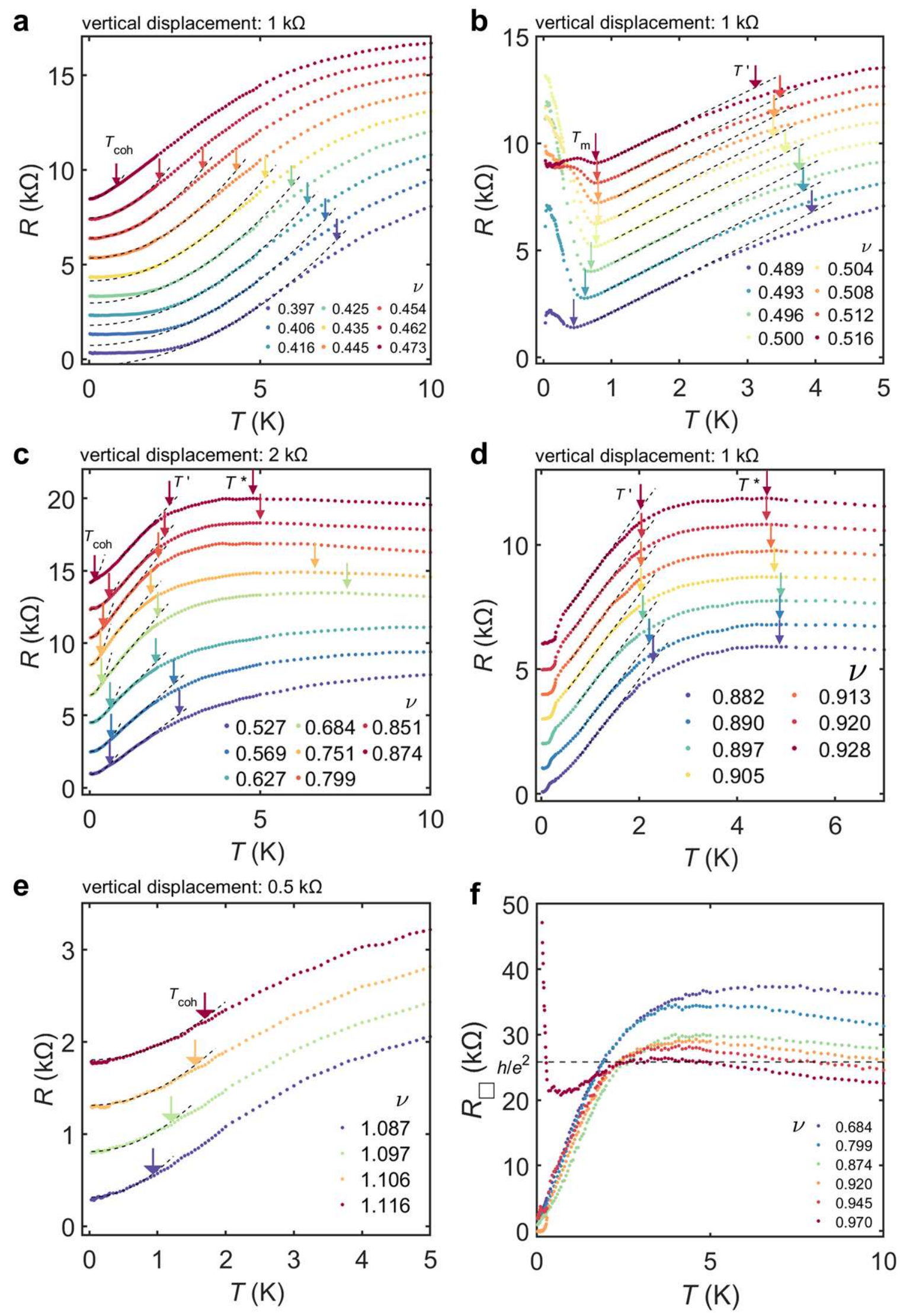

**Extended Data Fig. 6 | Temperature dependence analysis at $D$ = 0**

Temperature dependence of the resistance and extraction of the characteristic temperature in each region of **Fig. 3e**. **a,** Low-density region polarised to the middle layer. Across the entire region, the resistance decreases monotonically as the temperature is lowered within the measurement range, saturating at the residual resistance. $T_{coh}$ was estimated as the point where the quadratic fitting $R(T) = R_0 + A\ T^2$ deviated by 10%. **b**, Region near the $\nu$ = 1/2 correlated insulator. It exhibits insulating behaviour near the base temperature, but the resistance reaches a minimum at $T_m \approx$ 700 mK; subsequently, the resistance increases linearly as the temperature is raised, after which the rate of increase gradually decreases. $T'$ was estimated as the point where the data deviated by 10% from the linear fit. **c**, Region between the $\nu$ = 1/2 correlated insulator and SC1. Near the base temperature, the resistance exhibits a $T^2$ temperature dependence, which then becomes linear before gradually slowing down as the temperature rises. From $\nu$ > 0.63 onwards, the resistance begins to show a very broad peak at $T^*$, and decreases after the peak. **d**, SC1 region. A linear temperature dependence can be observed above the superconducting transition temperature; thereafter, the rate of increase decreases before reaching a maximum value. **e**, Region beyond the $\nu$ = 1 Mott insulator. A $T^2$ dependence is observed again at low temperatures, and the resistance increases monotonically within the measurement range. **f**, Sheet resistance $R_□$ calibrated by geometric factors. The horizontal dashed line marks the resistance quantum $h/e^2$.

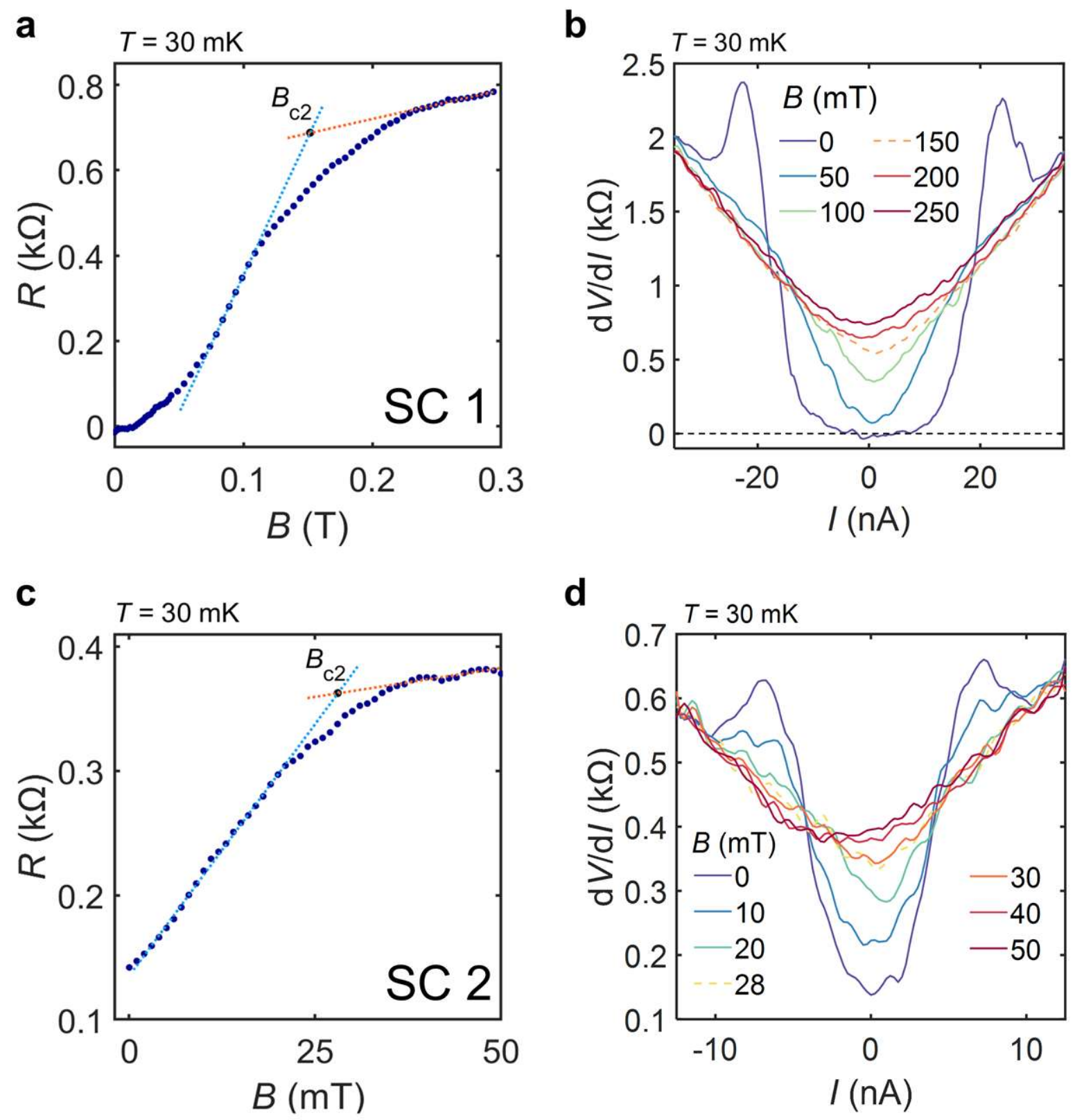

**Extended Data Fig. 7 | Estimation of the critical field $B_{c2}$**

**a**, Magnetic field dependence of the longitudinal resistance $R$ at 30 mK for SC1 ($\nu = 0.92$, $D = 0$). The critical field $B_{c2}$ is defined as the intersection point of the normal-state fit (orange line) and the unpinned-vortex-model fit (cyan line). **b**, Bias dependence of the differential resistance d$V$/d$I$ at various magnetic fields. The dotted line corresponds to the critical field $B_{c2}$. **c,d**, Measurements for SC2 ($\nu = 0.52$, $D = 122$ mV nm$^{-1}$) corresponding to **a** and **b**. Unlike SC1, $R$ does not reach zero, but the response of nonlinear transport to the magnetic field exhibits a similar pattern to that of SC1.

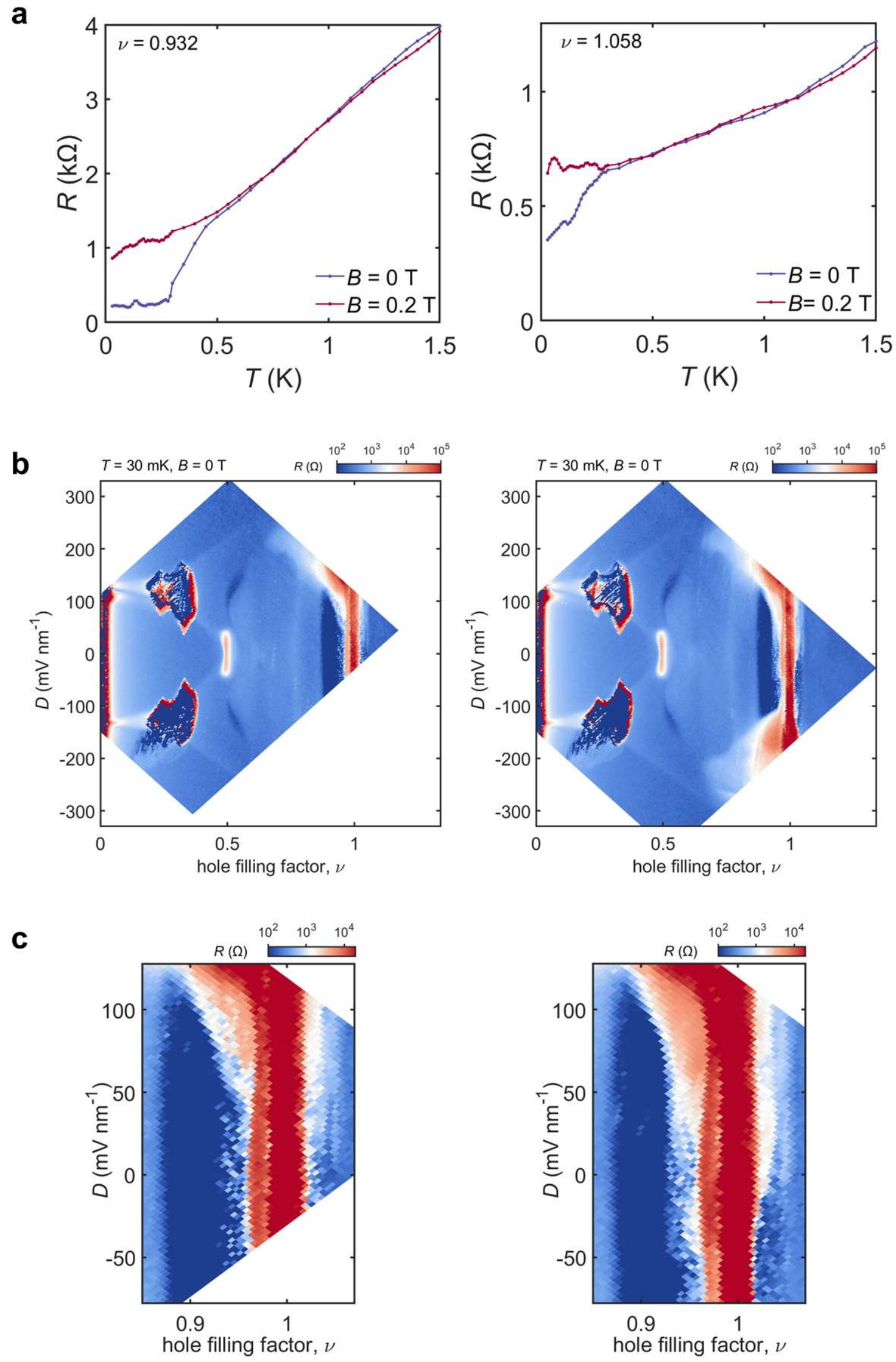

**Extended Data Fig. 8 | Resistance fluctuations in SC1**

**a**, Temperature dependence of the longitudinal resistance $R$ in the fluctuating SC1 region at $D = 0$ (left: $\nu = 0.932$, right: $\nu = 1.058$). Although the resistance does not reach zero on either side of the Mott insulator, a sharp drop in resistance is observed below a certain temperature, and this drop disappears when a magnetic field is applied. **b,** Two measurements of longitudinal resistance as a function of $\nu$ and $D$ at $T = 30$ mK and $B = 0$ T. The sample was thermally cycled to room temperature between the two measurements, and the behaviour of the fluctuating superconducting region differs. The resistance patterns are almost identical outside this region. **c**, Zoomed-in image of **b**.

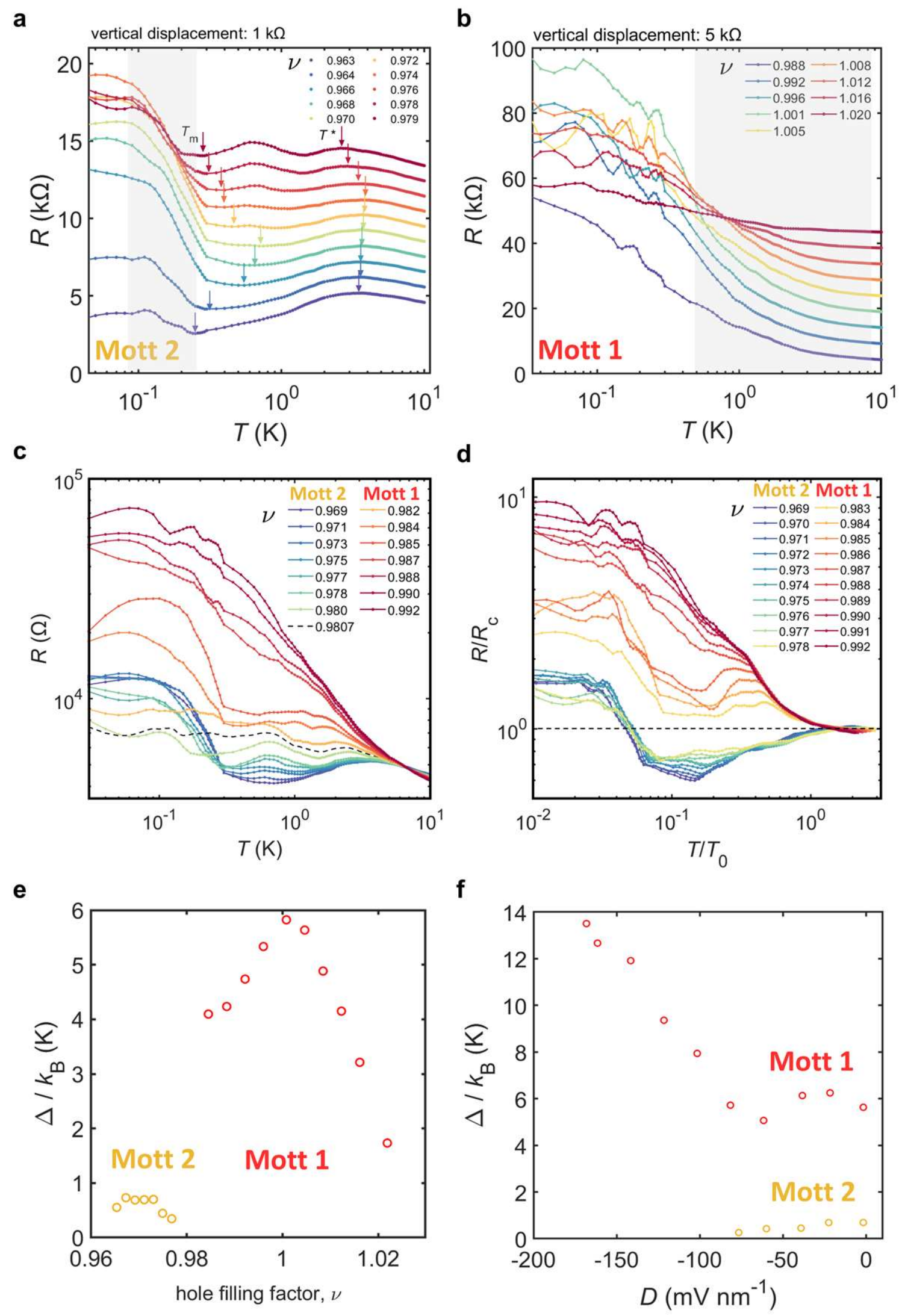

**Extended Data Fig. 9 | Insulator-to-insulator transition**

**a**, Temperature dependence of the longitudinal resistance $R$ at varying filling factor $\nu$ in the Mott2 region. Near the base temperature, insulating behaviour is observed; however, after $R$ reaches a minimum at $T_m$, the resistance increases as the temperature rises, and then decreases again after reaching a maximum at $T^*$. **b**, Temperature dependence of the longitudinal resistance $R$ at varying filling factor $\nu$ in the Mott1 region. $R$ decreases monotonically within the measurement range, exhibiting activation behaviour. The grey regions in **a** and **b** represent the temperature ranges for which the gap shown in **e** and **f** has been estimated using the Arrhenius fits. **c**, Temperature dependence of $R$ as a function of the filling factor $\nu$ near the Mott1–2 boundary. In the Mott2 region, as holes are doped, the resistance value $R_0$ at the base temperature (30 mK) decreases; after reaching a minimum at $\nu = 0.9807$, $R_0$ increases as the system approaches the Mott1 region, and it can be observed that the temperature dependence of the resistance is also weakest here. **d**, Temperature dependence of the normalised resistance. The resistance values at each $\nu$ were normalised by the resistance value at $\nu_c = 0.9807$, and the temperature was scaled to $T_0$, the temperature corresponding to the gap in the Mott1 region. In the Mott2 region, the $T_0$ value of the Mott1 state at the same distance from $\nu_c$ was used. While the Mott2 branch clearly exhibits a point where the minimum is reached, the Mott1 side is relatively monotonic. **e, f**, Comparison of the gap between the two insulating states as a function of doping (**e**) and $D$-field (**f**). Mott1 has a relatively large gap, and it can be observed that the gap tends to increase as the $D$-field is increased.

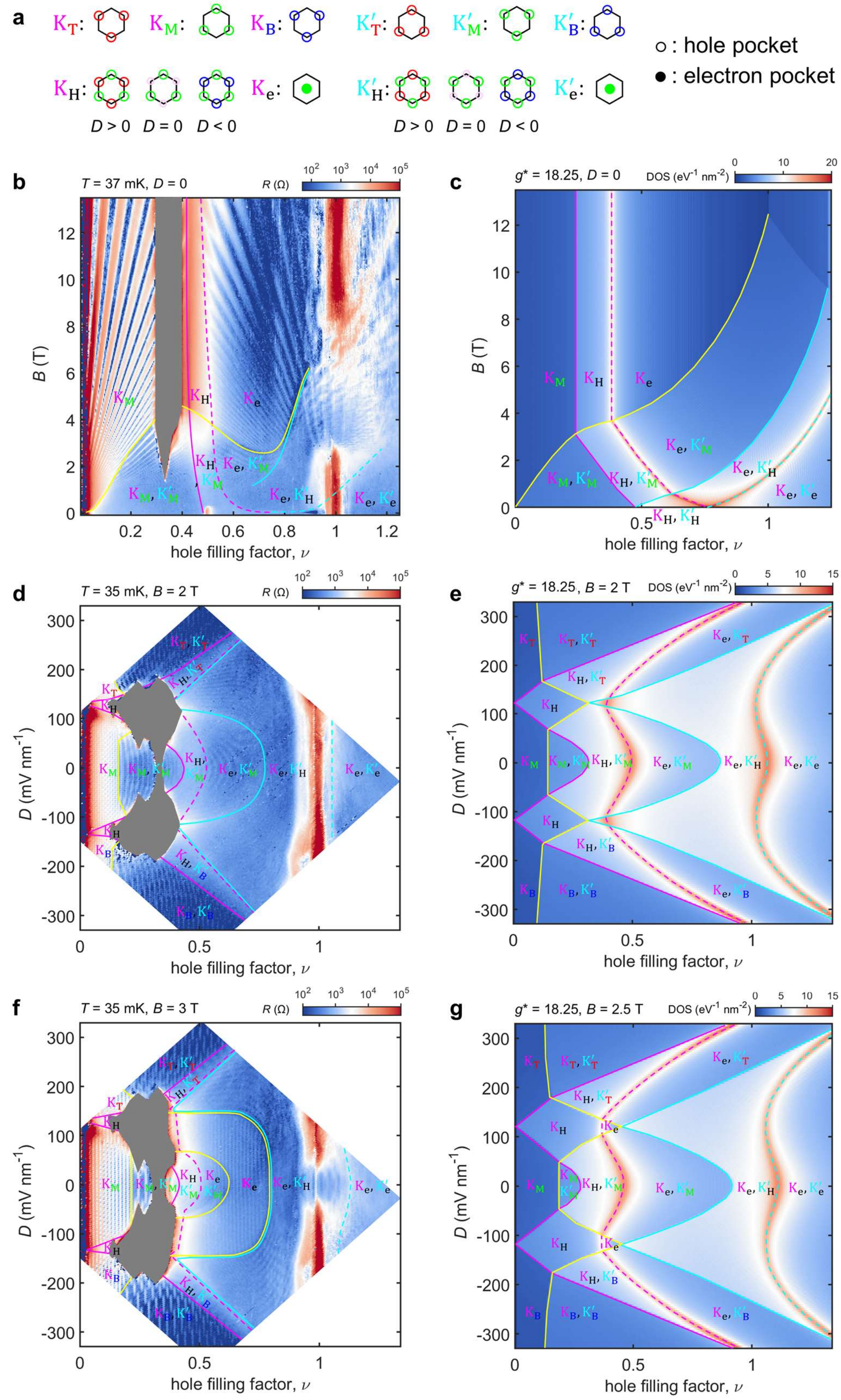
○ : hole pocket
● : electron pocket
D > 0 D = 0 D < 0
T = 37 mK, D = 0
g* = 18.25, D = 0
T = 35 mK, B = 2 T
g* = 18.25, B = 2 T
T = 35 mK, B = 3 T
g* = 18.25, B = 2.5 T
hole filling factor, ν
B (T)
D (mV nm-1)
R (Ω)
DOS (eV-1 nm-2)

**Extended Data Fig. 10 | Fermiology of $TTWSe_2$.**
**a**, Labels for the Fermi surface topologies used in **b–g**. Each colour corresponds to the same layer polarisation as in the main text. Empty and filled regions correspond to the hole and electron pockets, respectively. **b,c**, Longitudinal resistance (**b**, copy of **Extended Data Fig. 5g**) and the calculated DOS (**c**) as a function of $\nu$ and $B$ at $D = 0$. The effective g-factor was adjusted to fit the boundary between the $K_M$ region and the $K_M$, $K'_M$ region at low carrier density. **d,e,** Longitudinal resistance as a function of $\nu$ and $D$ (**d**, copy of **Extended Data Fig. 5d**) and the calculated DOS (**e**) at $B = 2$ T. **f,g**, Same as **d,e** at $B = 3$ T (DOS was calculated at 2.5 T for appropriate comparison with the data). In **b–g**, the yellow solid line separates the valley-polarised and valley-hybridised regions; the magenta (cyan) solid line separates the layer polarisation regions of the K (K′) valley bands; and the magenta (cyan) dashed line, representing vHS, separates the electron-like and hole-like regions of the K (K′) valley bands. In **c**, because the effective g-factor is fixed to be constant, the magnetic field at which valley polarisation occurs increases monotonically with $\nu$. However, the measured data in **b** show a downward bending in the layer-hybridised region, reaching a minimum (~ 2.5 T) near vHS ($\nu \approx 0.75$), indicating that the spin–valley susceptibility has been significantly enhanced. Below this magnetic field minimum, it can be seen that the $R(\nu, D)$ map (**d**) agrees qualitatively with the single-particle DOS calculation (**e**). In contrast, above this minimum, a C-shaped $K_e$ region, which is not expected in the DOS calculation (**g**), can be observed (**f**).

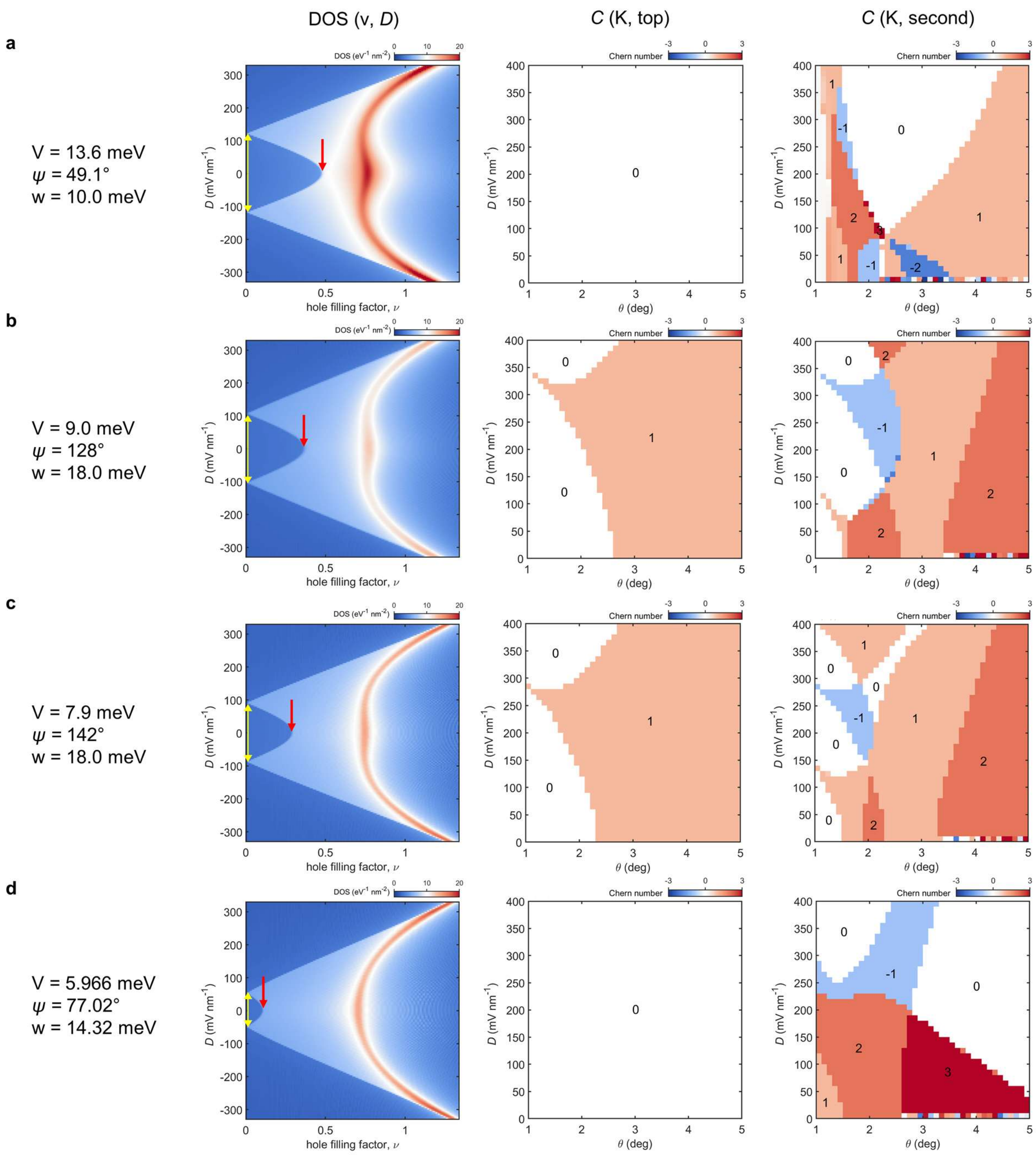

**Extended Data Fig. 11 | Continuum model calculation with various parameter sets**
**a**–**d**, Each row corresponds to the DOS ($\nu$, $D$), the Chern number of the topmost and second K-valley valence bands as a function of the twist angle $\theta$ and vertical displacement field $D$, calculated using continuum parameter values taken from the literature (**a**: ref 48, **b**: ref 4, **c**: ref 44, **d**: ref 11). The parameter set whose DOS ($\nu$, D) phase boundaries (yellow and red arrows) best match the experimental data (**a**) was used in the calculations presented in the main text. The numbers within the panels of the second and third columns represent the Chern number of each ($\theta$, $D$) region.

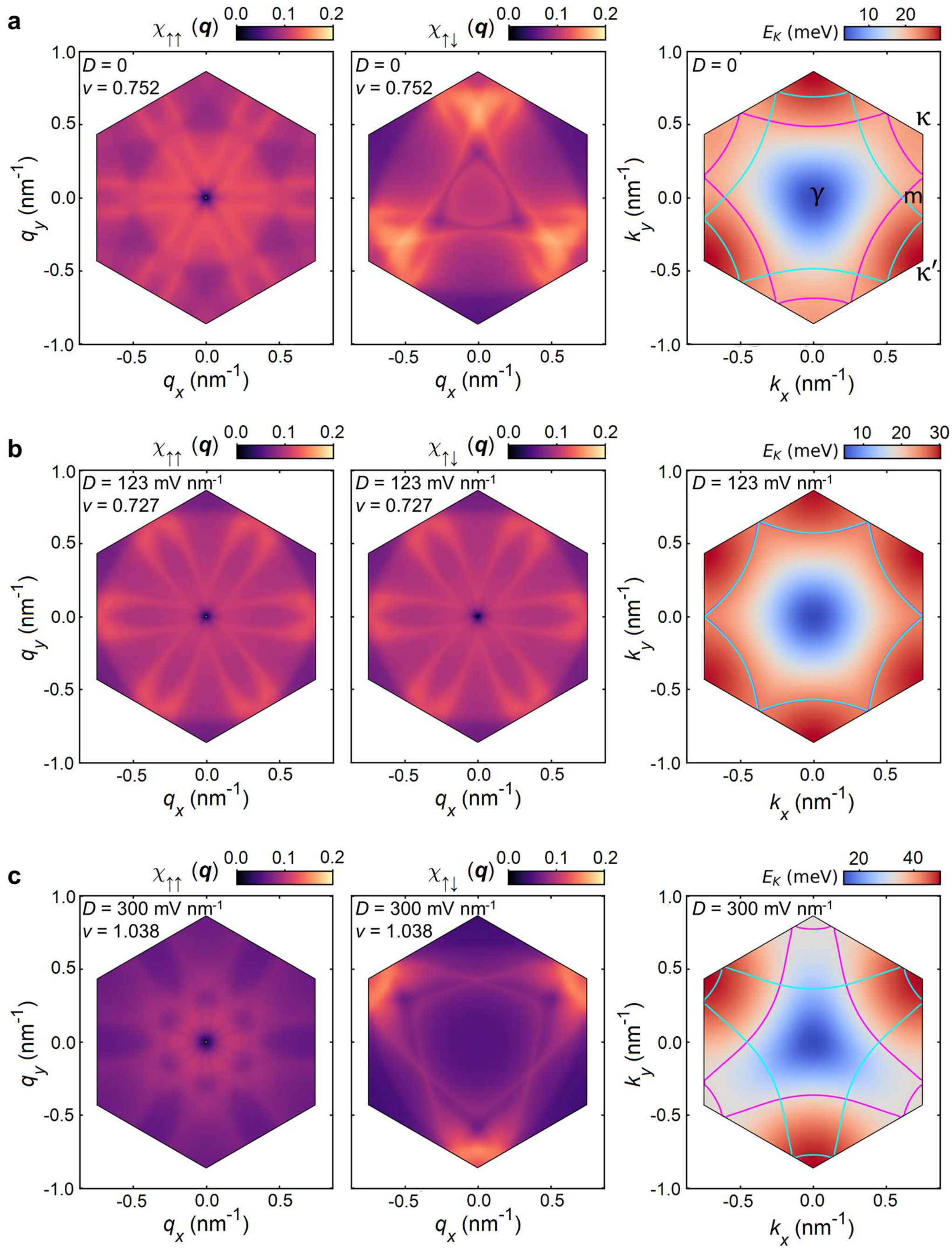

**Extended Data Fig. 12 | Particle–hole susceptibility at vHS filling**

Each columns represent the intravalley susceptibility $\chi_{\uparrow\uparrow}(\boldsymbol{q})$, intervalley susceptibility $\chi_{\uparrow\downarrow}(\boldsymbol{q})$ at vHS filling, and the band structure of the K-valley band. Magenta and cyan curves in third column represent the Fermi-surface contour of the K and K′-valley band at vHS filling, respectively. Each row corresponds to a different *D*-field. *D* = 0 (**a**), *D* = 123 mV nm$^{-1}$ (**b**), *D* = 300 mV nm$^{-1}$ (**c**). As *D* is increased, the position of the vHS shifts, and it can be seen that the point where intervalley susceptibility $\chi_{\uparrow\downarrow}(\boldsymbol{q})$ reaches its maximum moves from an incommensurate point m and then to κ.